\documentclass[twocolumn,showpacs,preprintnumbers,amsmath,amssymb]{revtex4-1}
\usepackage{graphicx} 
\usepackage{dcolumn} 
\usepackage{bm} 
\usepackage{braket} 
\usepackage{physics}
\usepackage[usenames,dvipsnames]{color}
\usepackage[
colorlinks=true, citecolor=blue, urlcolor=blue, linkcolor=blue,
setpagesize=false, bookmarks=false
]{hyperref}
\usepackage{ulem} 

\begin{document}


\title{Superconductivity and charge density wave under a time-dependent periodic field \\ in the one-dimensional attractive Hubbard model}
\author{Ryo Fujiuchi$^{1}$}
\author{Tatsuya Kaneko$^{2}$}
\author{Koudai Sugimoto$^{3}$}
\author{Seiji Yunoki$^{4,5,6}$}
\author{Yukinori Ohta$^{1}$}
\affiliation{
$^1$Department of Physics, Chiba University, Chiba 263-8522, Japan\\
$^2$Department of Physics, Columbia University, New York, New York 10027, USA\\
$^3$Department of Physics, Keio University, Yokohama 223-8522, Japan\\
$^4$Computational Condensed Matter Physics Laboratory, RIKEN Cluster for Pioneering Research (CPR), Wako, Saitama 351-0198, Japan\\
$^5$Computational Quantum Matter Research Team, RIKEN Center for Emergent Matter Science (CEMS), Wako, Saitama 351-0198, Japan\\
$^6$Computational Materials Science Research Team, RIKEN Center for Computational Science (R-CCS), Kobe, Hyogo 650-0047, Japan 
}
\date{\today}


\begin{abstract}  
We investigate the competition between superconductivity (SC) and charge density wave (CDW) under a time-dependent 
periodic field in the attractive Hubbard model. 
By employing the time-dependent exact diagonalization method, we show that 
the driving frequency and amplitude of the external field can control the enhancement of either the superconducting pair 
or the CDW correlation in the system, for which SC and CDW are degenerate in the ground state of the half-filled attractive Hubbard model in the absence of the field. 
In the strong-coupling limit of the attractive Hubbard interaction, the controllability is characterized by the anisotropic interaction 
of the effective model. The anisotropy is induced by the external field and lifts the degeneracy of SC and CDW.  
We find that the enhancement or suppression of the superconducting pair and CDW correlations in the periodically-driven 
attractive Hubbard model 
can be well interpreted by the quench dynamics of the effective model derived in the strong-coupling limit. 
\end{abstract}

\maketitle


\section{Introduction}
Field driven nonequilibrium systems have attracted much attention as a platform of new states of matter~\cite{zhang2014,basov2017,ishihara2019}.  
In these systems, light control and detection of intriguing electronic and structural properties are implemented 
by the ultrafast pump-probe spectroscopy~\cite{giannetti2016}.
One striking example of recent experimental observations is the light induced superconducting like properties
in some high-$T_c$ cuprates~\cite{fausti2011,hu2014,kaiser2014,nicoletti2014} and alkali-doped fullerides~\cite{mitrano2016,cantaluppi2018}, 
which have stimulated many theoretical investigations~\cite{sentef2016,patel2016,knap2016,kennes2017,sentef2017,babadi2017,murakami2017,mazza2017,wang2018}. 
On the other hand, quantum systems under a time-dependent periodic field are interpreted with the Floquet 
formalism~\cite{floquet1883}, which is also employed to design new quantum materials~\cite{oka2019}.

Here, we address how superconductivity (SC) and charge density wave (CDW) are influenced under a time-dependent periodic field. 
For this purpose, we consider the attractive Hubbard model at half-filling, which is a minimal model hosting SC and CDW as the 
ground state~\cite{micnas1990}, with a time-dependent periodic electric field introduced via the 
Peierls substitution~\cite{sentef2017a,kitamura2016}. 
In the weak-coupling regime of the attractive Hubbard interaction, 
the previous mean-field analysis reveals that CDW (SC) is enhanced (suppressed) when 
$\omega_p < 2\Delta_0$ (the field frequency $\omega_p$ is smaller than the single-particle energy gap $2\Delta_0$), 
while SC (CDW) is enhanced (suppressed) when $\omega_p>2\Delta_0$~\cite{sentef2017a}. 
In the strong-coupling regime, introducing the effective model for doublons, 
the strong-coupling expansion with the Floquet formalism has shown that $\eta$-pairing~\cite{yang1989} can possibly be induced 
due to the sign inversion of the pair hopping amplitude in the effective model~\cite{kitamura2016}.

In this paper, in order to explore  
the dynamics of the model in the entire driving regime, 
we employ the time-dependent exact diagonalization (ED) method, and 
we investigate the superconducting pair and CDW correlations 
in the periodically driven one-dimensional (1D) attractive Hubbard model at half-filling.  
We show how the superconducting pairing and CDW correlations are modified in a wide range of control parameters, 
including the field amplitude and frequency. 
When the external field is small, the behavior of the enhancement of SC and CDW shows good qualitative 
correspondence with the results in the weak-coupling mean-field analysis~\cite{sentef2017a}.  
With the strong attractive Hubbard interaction $U$, the CDW (superconducting pair) correlation is enhanced (suppressed) 
when $\omega_p < U$,  
while the superconducting pair (CDW) correlation is enhanced (suppressed) when $\omega_p >U$. 
We can interpret the mechanism on the basis of the anisotropic effective Heisenberg model derived by the strong-coupling expansion 
in the Floquet formalism.  
When the external field is strong, the modification of the superconducting pair and CDW correlations shows the complex 
parameter dependence, which is not simply interpreted by the ground-state phase diagram of the effective model in equilibrium. 
We find that these behaviors can be understood from the nonequilibrium dynamics after a quench of the effective interactions 
in the anisotropic effective Heisenberg model.

The rest of this paper is organized as follows. 
In Sec.~{\ref{sec:modelmethod}}, we introduce the model and briefly explain the method to study the time evolution of the pair and 
charge density correlations under the time-dependent periodic field.   
In Sec.~{\ref{sec:results}, we provide the numerical results for the attractive Hubbard model
,and we interpret these behaviors in terms of the equilibrium ground-state phase diagram of the strong-coupling effective model as well as 
the quench dynamics in the strong-coupling effective model. 
A summary is provided in Sec.~{\ref{sec:summary}}. 


\section{Model and Method}\label{sec:modelmethod}
\subsection{Attractive Hubbard model}\label{sec:ahm}

Here, we consider the 1D attractive Hubbard model
defined by the following Hamiltonian:
\begin{align}
  {\hat {\mathcal{H}}}  =  &- t_h\sum_{j=1}^{L}\sum_{\sigma}  \left( {\hat c}_{j,\sigma}^{\dag}{\hat c}_{j+1,\sigma} + {\rm H.c.} \right) - U  \sum_{j=1}^L  {\hat n}_{j, \uparrow}  {\hat n}_{j, \downarrow}, \label{eq:negU} 
\end{align}
where $\hat{c}_{j,\sigma}$($\hat{c}^\dagger_{j,\sigma}$) is the annihilation (creation) operator of an electron at site $j$ 
with spin $\sigma\, (=\uparrow,\downarrow)$, and $\hat{n}_{j,\sigma}=\hat{c}^\dagger_{j,\sigma}\hat{c}_{j,\sigma}$. 
$t_h$ is the hopping integral between the nearest-neighboring sites and $U$ ($>0$) is the on-site attractive interaction.
The number of sites $L$ is taken to be even, and we consider the half-filled case with the same number of up and down 
electrons, i.e., $N_{\uparrow}=N_{\downarrow}=L/2$.

In the strong-coupling limit $U \gg t_h$, 
up and down electrons tend to form an on-site pair, and no singly occupied sites are favored. 
Neglecting singly occupied sites,
the low-energy effective Hamiltonian $\hat{\mathcal{H}}_{\rm eff}$ in the strong-coupling limit is described by 
\begin{eqnarray}
  \hat{\mathcal{H}}_{\rm eff} &=& -\frac{J_0}{2}\sum_{j=1}^L \qty(\hat{c}^\dagger_{j,\downarrow}\hat{c}^\dagger_{j,\uparrow}\hat{c}_{j+1,\uparrow}\hat{c}_{j+1,\downarrow}+{\rm H.c.}) \nonumber \\ 
                              & +& V_0 \sum_{j=1}^L \hat{n}_{j,d} \hat{n}_{j+1,d} 
\label{eq:doublon} 
\end{eqnarray}
with $J_0=V_0=4t^2_h/U$, where $J_0$ is the pair hopping amplitude and $V_0$ is the nearest-neighbor pair 
repulsion~\cite{rosch2008,kitamura2016}. 
Here, $\hat{n}_{j,d}={\hat n}_{j, \uparrow}  {\hat n}_{j, \downarrow}$ is the number of doublons (doubly occupied electrons) at site $j$.

The effective Hamiltonian $\hat{\mathcal{H}}_{\rm eff}$ in Eq.~(\ref{eq:doublon}) can be expressed as the notion of pseudospin operators.
If the lattice is bipartite, one can define pseudospin operators via 
\begin{align}
  \begin{split}
    \hat{\eta}^+_j&=\hat{\eta}^x_j+i\hat{\eta}^y_j=(-1)^j\hat{c}^\dagger_{j,\downarrow}\hat{c}^\dagger_{j,\uparrow}, \\
    \hat{\eta}^-_j&=\hat{\eta}^x_j-i\hat{\eta}^y_j=(-1)^j\hat{c}_{j,\uparrow}\hat{c}_{j,\downarrow}, \\    
    \hat{\eta}^z_j&=\frac{1}{2}(\hat{n}_{j,\uparrow}+\hat{n}_{j,\downarrow}-1).
  \end{split}
\end{align}
These operators are called $\eta$-spin (or $\eta$-pairing) operators, which satisfy SU(2) algebra~\cite{yang1990,essler2005}. 
Note that $\hat{\eta}^z_j$ plays the same role with $\hat{n}_{j,d} -1/2$ when there is no singly occupied site in this strong-coupling model.  
It is easy to show that the effective Hamiltonian $\hat{\mathcal{H}}_{\rm eff}$ in Eq.~(\ref{eq:doublon}) can be mapped 
onto the isotropic (i.e., $J_0=V_0$) Heisenberg model with these $\eta$ operators: 
\begin{align}
  \hat{\mathcal{H}}_{\rm eff} &= J_0\sum_{j=1}^L\qty(\hat{\eta}_j^x\hat{\eta}_{j+1}^x+\hat{\eta}^y_j\hat{\eta}_{j+1}^y) + V_0 \sum_{j=1}^L\hat{\eta}_j^z\hat{\eta}_{j+1}^z. \label{eq:eta-spin}
\end{align} 
This pseudospin Hamiltonian 
is equivalent to the spin-$1/2$ isotropic Heisenberg Hamiltonian under the Shiba transformation~\cite{shiba1972,emery1976}. 
The $xy$ and $z$ components of the antiferromagnetism in this effective model correspond to the SC and CDW in the 
original attractive Hubbard model, respectively. 
They are degenerate because of the SU(2) symmetry ($J_0=V_0$).  

\subsection{External field}

The time-dependent external field is introduced in the hopping term in Eq.~(\ref{eq:negU}) 
via the Peierls substitution 
\begin{align}
t_h {\hat c}_{j,\sigma}^{\dag}{\hat c}_{j+1,\sigma} \; \rightarrow \; 
t_h e^{iA(t)}  {\hat c}_{j,\sigma}^{\dag}{\hat c}_{j+1,\sigma} , 
\label{eq:phase}
\end{align}
with the time-dependent vector potential $A(t)$.
Here, the velocity of light $c$, 
elementary charge $e$, Planck constant $\hbar$, and the lattice constant are all 
set to 1.  
In this paper, we consider the periodic driving external field given as 
\begin{align}
A(t) = 
\begin{cases}
  A_0 e^{-(t-t_0)^2/(2\sigma_p^2)} \cos \left[ \omega_p (t-t_0)  \right] & (t\leq t_0) \\
  A_0 \cos \qty[ \omega_p (t-t_0) ] & (t > t_0)
\end{cases}
\label{A(t)}
\end{align}
with the amplitude $A_0$ and frequency $\omega_p$.  
Corresponding to a semi-infinite ac field~\cite{ono2017}, this external field is introduced with the width $\sigma_p$ in time 
and becomes time-periodic for $t>t_0$.


\subsection{Method and correlation functions} \label{sec:method}
In the presence of the external field $A(t)$, the Hamiltonian is time-dependent, $\hat{\mathcal{H}}\rightarrow\hat{\mathcal{H}}(t)$, 
and hence we have to solve the time-dependent Schr$\ddot{\rm o}$dinger equation to evolve the state $\ket{\Psi(t)}$ in time. 
To obtain the exact time-evolved state for a long time ($t \le 300/t_h$)~\cite{DMRG},
we employ the time dependent ED method based on the Lanczos algorithm,  
where the time evolution with a short time step $\delta t$ is calculated  
in the corresponding Krylov subspace generated by $M_{\rm L}$ Lanczos iterations~\cite{Mohankumar2006,Park1986}. 
In our calculation, we use the finite-size clusters of $L$ sites with periodic boundary conditions (PBC). 
As the initial condition, we assume $\ket{\Psi(t=0)}=\ket{\psi_0}$, where $\ket{\psi_0}$ is the ground state of $\hat{\mathcal{H}}$ 
without the external field. 
We adopt $\delta t = 0.01/t_h$ and $M_{\rm L} = 15$ for the time evolution.  

In order to estimate the superconducting pair correlation, 
we calculate the time-dependent pair structure factor
\begin{align}
  P(q,t) = \dfrac{1}{L} \sum_{i,j} e^{iq\cdot (R_i-R_j)} \mel{\Psi(t)}{(\hat{\Delta}^\dag_{i}\hat{\Delta}_{j}+{\rm c.c.})}{\Psi(t)}, 
\end{align}
where $\hat{\Delta}_i = \hat{c}_{i,\uparrow}\hat{c}_{i,\downarrow}$ is the on-site pairing operator and $R_j$ is the position of site $j$.   
To estimate the CDW correlation, we calculate the charge structure factor 
\begin{align}
  C(q,t) = \dfrac{1}{L} \sum_{i,j} e^{iq\cdot (R_i-R_j)} \mel{\Psi(t)}{(\hat{\rho}_{i}-\rho)(\hat{\rho}_{j}-\rho)}{\Psi(t)}, 
\end{align}
where $\hat{\rho}_i = \hat{n}_{i,\uparrow}+\hat{n}_{i,\downarrow}$ is the charge density operator and $\rho$ is the average density, 
which is $1$ at half-filling. 
These correlation functions satisfy $P(q=0,t) = C(q=\pi,t)$ at $t=0$ since SC and CDW are degenerate in the ground (initial) state at half-filling.  
We also calculate the double occupancy 
  \begin{align}
    n_d(t) = \dfrac{1}{L}\sum_i\mel{\Psi(t)}{\hat{n}_{i,\uparrow}\hat{n}_{i,\downarrow}}{\Psi(t)}.
  \end{align}

We indicate the time-averaged value of a structure factor $F(q,t)$ (e.g., $P(q,t)$ and $C(q,t)$) as 
\begin{align}
\overline{F}(q)&=\frac{1}{t_f-t_i}\int^{t_f}_{t_i}dt F(q,t), 
\end{align}
where $t_i$ and $t_f$ are the lower and upper limit of the time average, respectively. 
In order to examine the enhancement or suppression of the superconducting pair and CDW correlations, 
we calculate the difference between the time averaged value and the initial value given by  
\begin{align}
\varDelta F(q) &= \overline{F}(q) - F(q,t=0).
\end{align}

\section{Results}\label{sec:results}
\subsection{Attractive Hubbard model}

\begin{figure}[t]
\centering
\includegraphics[width=\columnwidth]{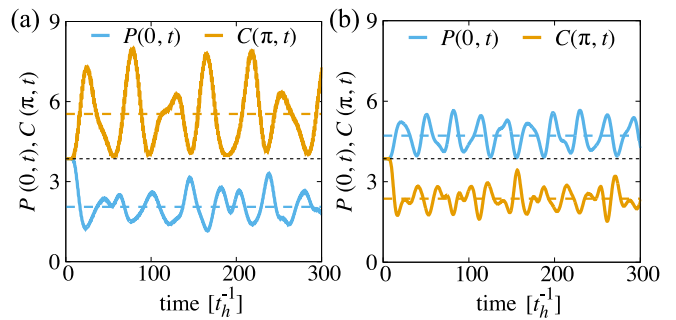}
\caption{Time evolution of the superconducting pair structure factor $P(q,t)$ at $q=0$ and the charge structure factor $C(q,t)$ 
at $q=\pi$ with (a) $\omega_p/U=0.15$ and $A_0 = 1$, and (b) $\omega_p/U=1.5$ and $A_0=1$. 
Dashed lines indicate $\overline{P}(q=0)$ (blue) and $\overline{C}(q=\pi)$ (orange) averaged from $t_i=0$ to $t_f=300/t_h$. 
Dotted black line indicates $P(q=0,t=0)$ and $C(q=\pi,t=0)$, which are degenerate in the initial state. 
The results are calculated by the ED method for $L=12$ (PBC) at $U=20t_h$ with $\sigma_p=2/t_h$ and $t_0=10/t_h$ in $A(t)$.
}
\label{fig:timedepend}
\end{figure}

\begin{figure}[t]
\centering
\includegraphics[width=\columnwidth]{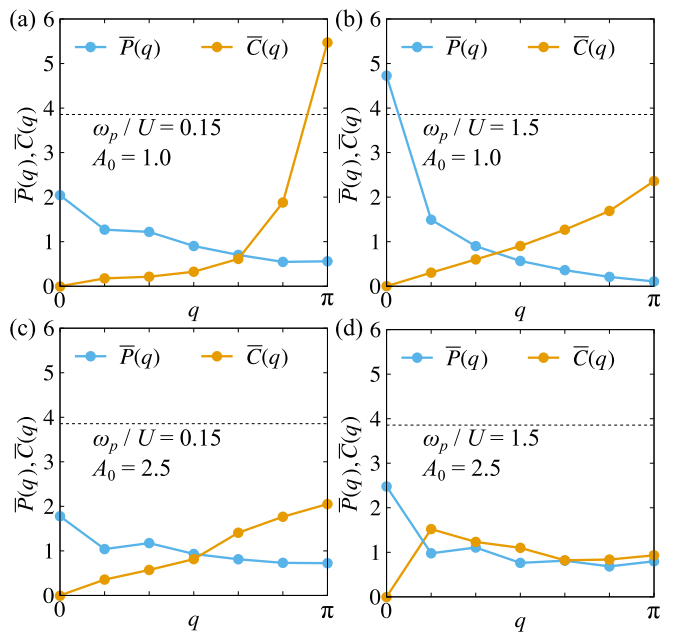}
\caption{Superconducting pair structure factor $\overline{P}(q)$ (blue) and charge structure factor $\overline{C}(q)$ (orange) 
averaged from $t_i=10/t_h$ to $t_f=100/t_h$ with (a) $\omega_p/U=0.15$ and $A_0 = 1$, (b) $\omega_p/U=1.5$ and $A_0=1$, 
(c) $\omega_p/U=0.15$ and $A_0 = 2.5$, and (d) $\omega_p/U=1.5$ and $A_0=2.5$.
Dotted line indicates $P(q=0,t=0)$ and $C(q=\pi,t=0)$, which are degenerate in the initial state.
The results are calculated by the ED method for $L=12$ (PBC) at $U=20t_h$ with $\sigma_p=2/t_h$ and $t_0=10/t_h$ in $A(t)$. 
}
\label{fig:qdepend}
\end{figure}

We first discuss the numerical results in the attractive Hubbard model. 
Figure~\ref{fig:timedepend} shows the time evolution of the superconducting pair correlation $P(q=0,t)$ and 
the CDW correlation $C(q=\pi,t)$. 
These structure factors $P(q=0,t)$ and $C(q=\pi,t)$ are indeed degenerate in the initial state at $t=0$. As shown in 
Fig.~\ref{fig:timedepend}(a), 
when the frequency $\omega_p$ is smaller than the attractive interaction, $\omega_p < U $, we find an enhancement of 
the CDW correlation $C(q=\pi,t)$ and a suppression of the superconducting pair correlation $P(q=0,t)$. 
In contrast, when $\omega_p > U $, $P(q=0,t)$ is enhanced, while $C(q=\pi,t)$ is suppressed, as compared to the initial value 
[see Fig.~\ref{fig:timedepend}(b)].
Although we take the large value of $U$ in Fig.~\ref{fig:timedepend}, these behaviors of the enhancement and suppression of 
the superconducting pair and CDW correlations are consistent with the results of the mean-field theory in 
the weak-coupling region~\cite{sentef2017a}. 

\begin{figure}[t]
\centering
\includegraphics[width=\columnwidth]{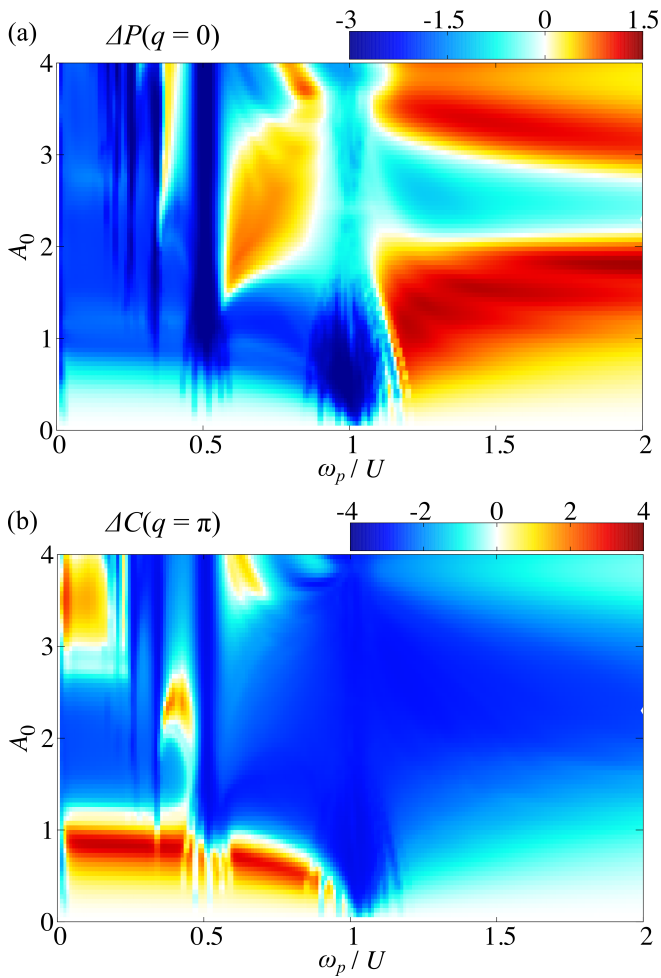}
\caption{Contour plots of 
  (a) the superconducting pair structure factor $\varDelta P(q=0)$ and 
  (b) the charge structure factor $\varDelta C(q=\pi)$ 
  in the parameter space of $\omega_p$ and $A_0$. 
$\varDelta P(q=0)$ and $\varDelta C(q=\pi)$ are averaged from $t_i=10/t_h$ to $t_f=100/t_h$. 
The results are calculated by the ED method for $L=12$ (PBC) at $U=20t_h$ with $\sigma_p=2/t_h$ and $t_0=10/t_h$ in $A(t)$.}
\label{fig:SCQ_NNQ}
\end{figure}

Figure~\ref{fig:qdepend} shows time-averaged $\overline{P}(q)$ and $\overline{C}(q)$ under the periodic driving field. 
As shown in Figs.~\ref{fig:qdepend}(a) and \ref{fig:qdepend}(b), when $A_0$ is small, 
$\overline{C}(q=\pi)$ is enhanced for $\omega_p<U$, while $\overline{P}(q=0)$ is enhanced for $\omega_p>U$, 
corresponding to the results in Fig.~\ref{fig:timedepend}. 
On the other hand, when $A_0$ is relatively large, e.g., $A_0=2.5$ in Figs.~\ref{fig:qdepend}(c) and \ref{fig:qdepend}(d), 
$\overline{P}(q=0)$ and $\overline{C}(q=\pi)$ are both suppressed from the initial value at $t=0$. 
It is also observed in Fig.~\ref{fig:qdepend} that, while the $\eta$-pairing correlation $P(q=\pi,t)$ is strongly enhanced by the optical 
pulse in the case of the repulsive model~\cite{kaneko2019,fujiuchi2019,kaneko2020}, $P(q,t)$ does not exhibit a sharp peak 
at $q=\pi$ in the attractive model with the periodic driving field $A(t)$ in Eq.~(\ref{A(t)}).

In order to explore the parameter dependence of the superconducting pair and CDW correlations,  
Fig.~\ref{fig:SCQ_NNQ} shows $\varDelta P(q=0)$ and $\varDelta C(q=\pi)$ with different values of $A_0$ and $\omega_p$. 
In the small $A_0$ ($\alt 1$) region, 
the CDW correlation $\overline{C}(q=\pi)$ is enhanced for $\omega_p < U$, while the superconducting pair correlation 
$\overline{P}(q=0)$ is enhanced for $\omega_p > U$. These results are in good qualitative agreement with the previous study 
using the mean-field theory~\cite{sentef2017a}. 
However, in the large-$A_0$ region, the parameter dependence of these correlations is not simple.  
For example, in the region around $2< A_0 < 3$, the superconducting pair correlation is suppressed even for $\omega_p > U$ 
but it is enhanced for $U/2 < \omega_p  < U$ [see Fig.~\ref{fig:SCQ_NNQ}(a)]. This behavior is opposite to the results found 
in the small-$A_0$ region. 
This complex behavior in the large-$A_0$ region is not simply interpreted by the mean-field picture with 
a small external field~\cite{sentef2017a}. 

We also notice in Fig.~\ref{fig:SCQ_NNQ} that the correlation functions 
around the parameters at $\omega_p = U/m$ ($m$: integer) are rather steeply suppressed. 
To understand this feature, we calculate the time evolution of the double occupancy $n_d(t)$.
As shown in Fig.~\ref{fig:nd}, the double occupancy $n_d(t)$ is strongly suppressed by the periodic field when 
$\omega_p = U/m$.   
This causes the steep suppressions of the correlation functions found in Fig.~\ref{fig:SCQ_NNQ}. 
On the other hand, 
the periodic field with $\omega_p $ away from $ U/m$ does not suppress the double occupancy $n_d(t)$. 
This is understood because in the off-resonant case~\cite{dunlap1986,messer2018,sandholzer2020}, 
$U/t^{\rm eff}_h$ characterized by the effective hopping $t^{\rm eff}_h=t_h \mathcal{J}_0(A_0)\,  < t_h$  ($\mathcal{J}_0(x)$: zeroth Bessel function) becomes  larger than the initial value  $U/t_h$.} 
Indeed, the double occupancy at an off-resonant frequency, e.g., $\omega_p/U=1.5$, is slightly enhanced 
from the initial value [see Fig.~\ref{fig:nd}(b)], which is suitable for the enhancement of the superconducting pair and CDW correlations. 

\begin{figure}[t]
\centering
\includegraphics[width=\columnwidth]{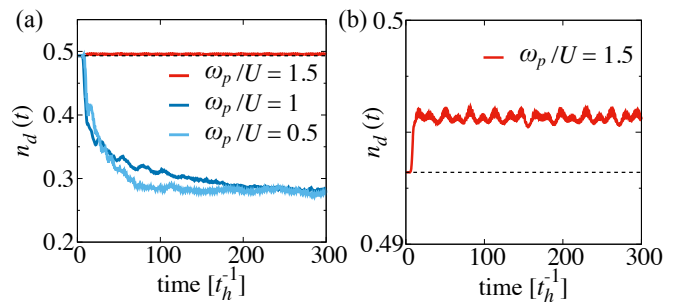}
\caption{
(a) Time evolution of the double occupancy $n_d(t)$ for $\omega_p/U=0.5, 1$, and $1.5$. 
(b) Enlarged plot of (a) for $\omega_p/U=1.5$ near $n_d(t) = 0.5$.
Dotted line indicates $n_d(t=0)$.
The results are calculated by the ED method for $L=12$ (PBC) at $U=20t_h$ with  $A_0 = 1$, $\sigma_p=2/t_h$ and $t_0=10/t_h$ in $A(t)$.
}
\label{fig:nd}
\end{figure}

\subsection{Effective model in the strong-coupling limit} \label{sec:effective}

To interpret the behavior of $P(q=0,t)$ and $C(q=\pi,t)$ in the wide parameter space, 
we now introduce the effective model derived by the strong-coupling expansion in the Floquet formalism~\cite{kitamura2016}. 
The strong-coupling expansion is expected to be valid away from $\omega_p = U/m$, 
where the double occupancy keeps $n_d(t)\sim 0.5$ (see Fig.~\ref{fig:nd}) and thus no singly occupied sites are favored 
in the time-evolved state.
Under the periodic driving field $A(t)=A_0 \cos\omega_p t$, the effective model for the attractive Hubbard model with a large 
$U$ is given by  
\begin{align}
 \hat{\mathcal{H}}_{\rm eff} &= J_{\rm eff}\sum_{j=1}^L\qty(\hat{\eta}_j^x\hat{\eta}_{j+1}^x+\hat{\eta}^y_j\hat{\eta}_{j+1}^y) + V_{\rm eff} \sum_{j=1}^L\hat{\eta}_j^z\hat{\eta}_{j+1}^z, \label{eq:eff-eta} 
\end{align}
with the effective interactions 
\begin{align}
 J_{\rm eff}&=\sum_{m=-\infty}^\infty(-1)^m\dfrac{4t_h^2\mathcal{J}_m(A_0)^2}{U+m\omega_p}, \label{eq:eff-J} \\ 
 V_{\rm eff}&= \sum_{m=-\infty}^\infty\dfrac{4t_h^2\mathcal{J}_m(A_0)^2}{U+m\omega_p}, \label{eq:eff-V}
\end{align}
where $\mathcal{J}_m(x)$ is the $m$th Bessel function~\cite{kitamura2016}.  
Notice that this effective model corresponds to an anisotropic Heisenberg (XXZ) model, and the effective interactions 
$J_{\rm eff}$ and $V_{\rm eff}$ vary in different manners, 
which is the manifestation of the broken $\eta$-SU(2) symmetry due to the external field $A(t)$. 
Therefore, the degeneracy of SC and CDW is lifted by the external field $A(t)$, and the anisotropy of $J_{\rm eff}$ and $V_{\rm eff}$ 
gives rise to the enhancement or suppression of the superconducting pair and CDW correlations. 
This should be contrasted with the strong-coupling expansion in the repulsive Hubbard model, 
for which the effective model is spin SU(2) symmetric (i.e., isotropic for the spin degrees of freedom) 
even in the presence of a time-dependent periodic electric field~\cite{mentink2015}.
As shown in Eqs.~(\ref{eq:eff-J}) and (\ref{eq:eff-V}), $J_{\rm eff}$ and $V_{\rm eff}$ diverge at $\omega_p = U/m$, 
also indicating that this strong-coupling expansion is not valid at $\omega_p = U/m$.

In the small-$A_0$ region, the enhancement or suppression of the superconducting pair and CDW correlations can be  
understood by the anisotropic effective interactions $J_{\rm eff}$ and $V_{\rm eff}$. 
When $A_0 \ll 1$, $J_{\rm eff}$ and $V_{\rm eff}$ are given by 
\begin{align}
 J_{\rm eff}&\approx \dfrac{4t_h^2}{U} \left(1- \dfrac{A_0^2}{2}\right) + \dfrac{2U t_h^2 }{\omega_p^2-U^2} A_0^2,  
 \\ 
 V_{\rm eff}&\approx \dfrac{4t_h^2}{U}\left(1- \dfrac{A_0^2}{2}\right) - \dfrac{2U t_h^2 }{\omega_p^2-U^2} A_0^2, 
\end{align}
Therefore, when $\omega_p > U$, $J_{\rm eff} > V_{\rm eff}$ and thus the superconducting pair correlation is enhanced, while 
when $\omega_p < U$, $V_{\rm eff} > J_{\rm eff}$ and hence the CDW correlation is enhanced.

However, in the large-$A_0$ region, the enhancement or suppression of $\varDelta{P}(q=0)$ and $\varDelta{C}(q=\pi)$ 
in Fig.~\ref{fig:SCQ_NNQ} is not simply interpreted by the ground-state phase diagram of the effective 
model $\hat{\mathcal{H}}_{\rm eff}$ in Eq.~(\ref{eq:eff-eta}). 
For instance, although $\eta$-pairing is anticipated when $J_{\rm eff}<0$ in the ground state of the effective model, 
$P(q,t)$ does not show a sharp peak at $q=\pi$ in the corresponding region [see, e.g., Fig.~\ref{fig:qdepend}(c)].  
This is because the time-evolved state under the external field $A(t)$ retains the memory of the initial state $\ket{\psi_0}$, 
and the system may not necessarily relax to the ground state of the effective model. 
This may be interpreted by the dynamical instability of the effective Hamiltonian discussed in Ref.~\cite{kitamura2016}.
Therefore, as shown below, the memory effect of the initial state has to be incorporated 
to understand the behavior of $P(q=0,t)$ and $C(q=\pi,t)$ in the wide parameter region.

\subsection{Quench dynamics of the effective model} \label{sec:quench}

\begin{figure}[t]
\centering
\includegraphics[width=\columnwidth]{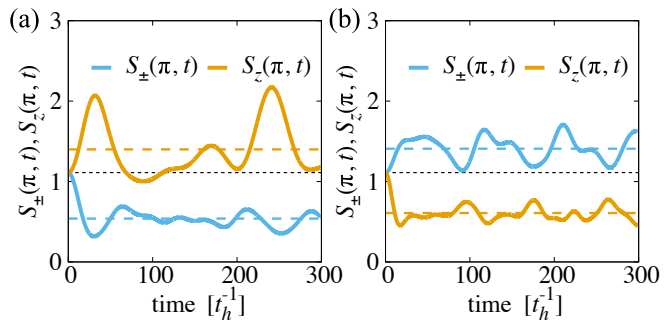}
\caption{Time evolution of the $xy$ and the $z$ components of the $\eta$-spin correlation functions, 
$S_{\pm}(q=\pi,t)$ and $S_{z}(q=\pi,t)$, respectively, 
with (a) $\omega_p/U=0.15$ and $A_0 = 1$, and (b) $\omega_p/U=1.5$ and $A_0=1$. 
We assume $J_{\rm eff}$ and $V_{\rm eff}$ at $U=20t_h$.
Dashed lines indicate $\overline{S}_{\pm}(q=\pi)$ (blue) and $\overline{S}_z(q=\pi)$ (orange) averaged from $t_i=0$ to $t_f=300/t_h$. 
Dotted black line indicates $S_{\pm}(q=\pi,t=0)$ and $S_z(q=\pi,t=0)$, which are degenerate in the initial state.
The results are calculated in the anisotropic Heisenberg (XXZ) model for $L=18$ (PBC). 
}
\label{fig:time_XXZ}
\end{figure}

\begin{figure}[t]
\centering
\includegraphics[width=\columnwidth]{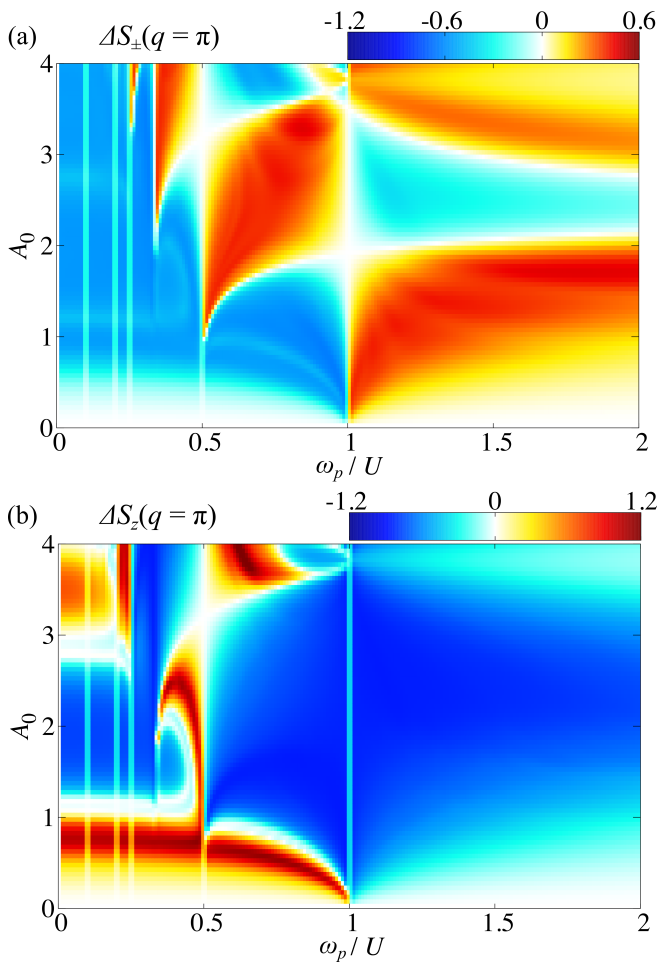}
\caption{Contour plots of
  (a) the $xy$-component of the $\eta$-spin correlation function  $\varDelta{S}_{\pm}(q=\pi)$ and
  (b) the $z$-component of the $\eta$-spin correlation function $\varDelta{S}_z(q=\pi)$ 
  after the parameter quench $(J_0,V_0) \rightarrow (J_{\rm eff},V_{\rm eff})$ in the parameter space of $\omega_p$ and $A_0$.
$\varDelta{S}_{\pm}(q=\pi)$ and $\varDelta{S}_z(q=\pi)$ are averaged from $t_i=0$ to $t_f=100/t_h$. 
We assume $J_{\rm eff}$ and $V_{\rm eff}$ at $U=20t_h$.
The results are calculated in the anisotropic Heisenberg (XXZ) model for $L=18$ (PBC). 
}
\label{fig:XXZ}
\end{figure}

To address this issue described above, here we investigate the nonequilibrium dynamics after a quench of the exchange coupling 
in the XXZ model $\hat{\mathcal{H}}_{\rm eff}$ in Eq.~(\ref{eq:eff-eta}).   
We set as the initial state the ground state of the isotropic Heisenberg model with $J_0=V_0$ in Eq.~(\ref{eq:eta-spin}), 
and we change the parameters 
to the effective values $J_{\rm eff}$ and $V_{\rm eff}$, given in Eqs.~(\ref{eq:eff-J}) and (\ref{eq:eff-V}), abruptly at time $t=0$. 
To examine the quench dynamics in the XXZ model, we calculate the time evolution of the $xy$ and $z$ components of 
the $\eta$-spin structure factors 
\begin{align}
S_{\pm}(q,t)&=\dfrac{1}{L}\sum_{i,j}e^{iq\cdot(R_i-R_j)} \mel{\Psi(t)}{\hat{\eta}_i^{+}\hat{\eta}_j^{-}+\hat{\eta}_i^{-}\hat{\eta}_j^{+}}{\Psi(t)}, \\
S_{z}(q,t)&=\dfrac{4}{L}\sum_{i,j}e^{iq\cdot(R_i-R_j)} \mel{\Psi(t)}{\hat{\eta}_i^{z}\hat{\eta}_j^{z}}{\Psi(t)},
\end{align}
corresponding to the pair and charge structure factors $P(q,t)$ and $C(q,t)$ in the attractive Hubbard model, respectively. 
Note that the $xy$ component of antiferromagnetic correlation $S_{\pm}(q=\pi,t)$ in the XXZ model corresponds 
to the superconducting pair correlation $P(q=0,t)$ in the attractive Hubbard model.

Figure~\ref{fig:time_XXZ} shows the time evolution of the $xy$ and $z$ components of the $\eta$-spin 
correlations, $S_{\pm}(q=\pi,t)$ and $S_{z}(q=\pi,t)$, respectively,  
after the parameter quench $(J_0,V_0) \rightarrow (J_{\rm eff},V_{\rm eff})$ in the small $A_0$ region.
The characteristic behavior of 
these correlation functions is in good agreement with the time evolution of $P(q=0,t)$ and $C(q=\pi,t)$ shown in Fig.~\ref{fig:timedepend}. 
The $z$ component of the $\eta$-spin correlation $S_z(q=\pi,t)$ is enhanced when $\omega_p<U$ (i.e., $V_{\rm eff} > J_{\rm eff}$), 
while the $xy$ component of the $\eta$-spin correlation $S_{\pm}(q=\pi,t)$ is enhanced when $\omega_p>U$ (i.e., $J_{\rm eff} > V_{\rm eff}$). 
Figure~\ref{fig:XXZ} shows the contour plots of $\varDelta{S}_{\pm}(q=\pi)$ and $\varDelta{S}_z(q=\pi)$ after the parameter quench 
in the wide parameter region of $A_0$ and $\omega_p$.
Figure~\ref{fig:XXZ} is in excellent qualitative agreement with $\varDelta{P}(q=0)$ and $\varDelta{C}(q=\pi)$ 
shown in Fig.~\ref{fig:SCQ_NNQ}, including the large $A_0$ region. 
Therefore, the quench dynamics of the effective XXZ model provides a good understanding of the behavior of the 
superconducting pair and CDW correlations in the original attractive Hubbard model under the periodic driving field.

 \subsection{Phase diagram} \label{sec:phase}
  
\begin{figure}[t]
\centering
\includegraphics[width=\columnwidth]{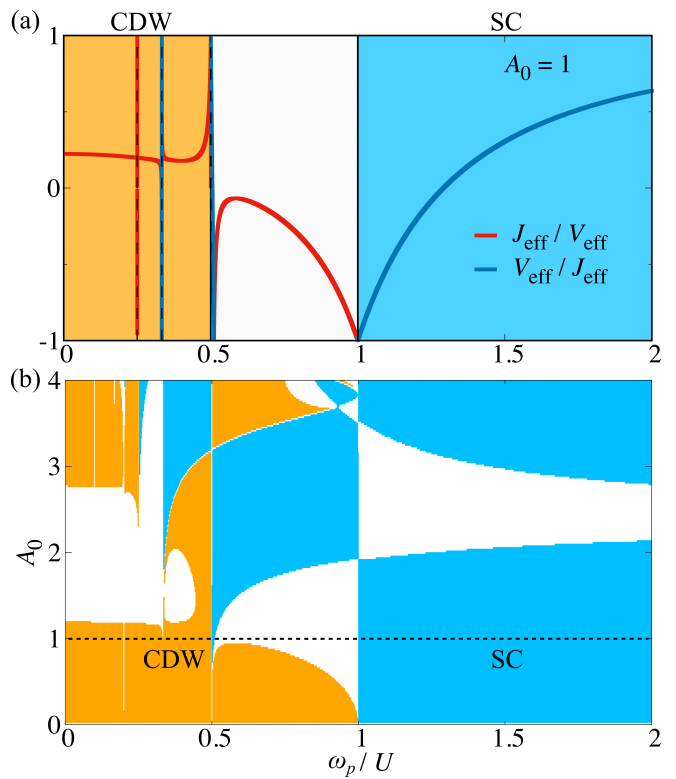}
\caption{
(a) The frequency dependence of the ratio of the effective coupling $J_{\rm eff}/V_{\rm eff}$ ($V_{\rm eff}/J_{\rm eff}$) 
with $A_0=1$.
(b) Phase diagram of the periodically-driven attractive Hubbard model in the strong-coupling regime at half-filling. 
SC is enhanced when $|J_{\rm eff}|>|V_{\rm eff}|$ (blue regions) and CDW is enhanced when $|V_{\rm eff}|>|J_{\rm eff}|$ 
with $V_{\rm eff}J_{\rm eff} > 0$ (orange regions).  
SC and CDW are both suppressed when $|V_{\rm eff}|>|J_{\rm eff}|$ with $V_{\rm eff}J_{\rm eff} < 0$ (white regions). 
}
\label{fig:analysis}
\end{figure}  

Finally, we summarize our finding by showing the phase diagram in Fig.~\ref{fig:analysis}(b) that can reproduce almost quantitatively 
the results of the enhancement or suppression of the superconducting pair and CDW correlations 
found in Fig.~\ref{fig:XXZ}. 
The phase diagram in Fig.~\ref{fig:analysis}(b) is constructed simply as follows: the SC is enhanced when $|J_{\rm eff}|>|V_{\rm eff}|$, 
the CDW is enhanced when $|V_{\rm eff}|>|J_{\rm eff}|$ with $V_{\rm eff}J_{\rm eff} > 0$, and  
the SC and CDW are both suppressed when $|V_{\rm eff}|>|J_{\rm eff}|$ with $V_{\rm eff}J_{\rm eff} < 0$. 
It should be emphasized that this phase diagram is determined from the effective interactions $J_{\rm eff}$ and $V_{\rm eff}$
of the effective model in 
Eq.~(\ref{eq:eff-eta}), not the ground-state phase diagram of the effective model, as demonstrated in Fig.~\ref{fig:analysis}(a) for the 
case of $A_0=1$.  

As in the ground state of the effective model, when $A_0$ is small, the SC (CDW) is enhanced (suppressed) in the region 
where $J_{\rm eff}$ dominates $V_{\rm eff}$ (i.e., $\omega_p > U$), and the CDW (SC) is enhanced (suppressed) 
in the region where $V_{\rm eff}$ dominates $J_{\rm eff}$ (i.e., $\omega_p < U$). In the large $A_0$ region, 
the $\eta$-pairing and the phase separation would be anticipated by considering the ground-state phase diagram of 
the effective model, where the former is favored when $J_{\rm eff}< 0$ and $|J_{\rm eff}|>|V_{\rm eff}|$, and the latter is favored 
when $V_{\rm eff}<0$ and $|V_{\rm eff}|>|J_{\rm eff}|$.
However, the tendency toward these 
is not observed in our calculations [see, e.g., Figs.~\ref{fig:qdepend}(c) and \ref{fig:qdepend}(d)]. 
As discussed in Sec.~\ref{sec:quench}, 
this is because the steady state driven by the periodic field retains the memory of the initial state, which can be 
captured rather well by the quench dynamics in the effective model. 
Including all these features, the phase diagram simply constructed in Fig.~\ref{fig:analysis}(b) is in excellent agreement 
with Figs.~\ref{fig:SCQ_NNQ} and ~\ref{fig:XXZ} in the wide parameter region of $\omega_p$ and $A_0$.


\section{Conclusion} \label{sec:summary}
We have investigated the change of the superconducting pair and charge correlations in the 1D periodically-driven attractive 
Hubbard model in the strong-coupling regime. 
When the external field is small, the CDW (superconducting pair) correlation is enhanced (suppressed) for $\omega_p < U$, 
while the superconducting pair (CDW) correlation is enhanced (suppressed) for $\omega_p >U$.  
This mechanism is well interpreted on the basis of the ground-state phase diagram of 
the effective anisotropic Heisenberg (XXZ) model derived by the strong-coupling expansion in the Floquet formalism, 
where the periodic driving field changes the effective interactions. 
When the external field is strong, the parameter dependence of the enhancement or suppression of the correlations is 
more complex and is not simply interpreted by the ground-state phase diagram of the effective model. 
We have shown that these behaviors can be understood from the nonequilibrium dynamics after a quench of the effective interactions 
in the effective model.  

We should note that the effective model studied here was originally introduced in Ref.~\cite{kitamura2016}. 
In their study, they mainly focused on the $\eta$-pairing that can be induced by a driving field~\cite{kitamura2016}, 
while here we have investigated 
the enhancement or suppression of the superconducting pair and CDW correlations by employing the unbiased ED method. 
However, we have confirmed that the $\eta$-pairing correlation can be induced even in the ED calculation when we adopt 
the specific protocol for the $\eta$-pairing, i.e., adiabatic change of the amplitude $A_0$, used in  Ref.~\cite{kitamura2016}. 

In the experimental side, the attractive Hubbard model is realized using an atomic Fermi gas in an optical lattice~\cite{mitra2018}, 
and the Floquet dynamics in the attractive model demonstrated here would be observable in the periodically driven Fermi-Hubbard system~\cite{messer2018,sandholzer2020}. 
The competition between SC and CDW has been observed in some cuprates~\cite{ghiringhelli2012,chang2012} and 
transition metal dichalcogenides~\cite{harper1977,neto2001}. 
The suppression of a competing order against SC is expected to play a key role in the light-induced  SC~\cite{fausti2011,nicoletti2014}.  
While we have considered a generic minimal model for SC and CDW, the effective attraction in the Holstein model becomes 
a Hubbard type~\cite{freericks1993} and therefore some of our finding might also be applied to electron-phonon systems. 


\begin{acknowledgments}
The authors acknowledge H. Aoki, S. Miyakoshi, Y. Murakami, K. Seki, and T. Shirakawa for fruitful discussion. 
This work was supported in part by Grants-in-Aid for Scientific Research from JSPS 
(Projects No.~JP17K05530, No.~JP18H01183, No.~JP18K13509, No.~JP19J20768, No.~JP19K14644, 
and No.~JP20H01849) of Japan 
and Keio University Academic Development Funds for Individual Research.
R.F. acknowledges support from the JSPS Research Fellowship for Young Scientists.
T.K. was supported by the JSPS Overseas Research Fellowship.
\end{acknowledgments}


\bibliography{bibliography}

\begin{thebibliography}{48}%
\makeatletter
\providecommand \@ifxundefined [1]{%
 \@ifx{#1\undefined}
}%
\providecommand \@ifnum [1]{%
 \ifnum #1\expandafter \@firstoftwo
 \else \expandafter \@secondoftwo
 \fi
}%
\providecommand \@ifx [1]{%
 \ifx #1\expandafter \@firstoftwo
 \else \expandafter \@secondoftwo
 \fi
}%
\providecommand \natexlab [1]{#1}%
\providecommand \enquote  [1]{``#1''}%
\providecommand \bibnamefont  [1]{#1}%
\providecommand \bibfnamefont [1]{#1}%
\providecommand \citenamefont [1]{#1}%
\providecommand \href@noop [0]{\@secondoftwo}%
\providecommand \href [0]{\begingroup \@sanitize@url \@href}%
\providecommand \@href[1]{\@@startlink{#1}\@@href}%
\providecommand \@@href[1]{\endgroup#1\@@endlink}%
\providecommand \@sanitize@url [0]{\catcode `\\12\catcode `\$12\catcode
  `\&12\catcode `\#12\catcode `\^12\catcode `\_12\catcode `\%12\relax}%
\providecommand \@@startlink[1]{}%
\providecommand \@@endlink[0]{}%
\providecommand \url  [0]{\begingroup\@sanitize@url \@url }%
\providecommand \@url [1]{\endgroup\@href {#1}{\urlprefix }}%
\providecommand \urlprefix  [0]{URL }%
\providecommand \Eprint [0]{\href }%
\providecommand \doibase [0]{http://dx.doi.org/}%
\providecommand \selectlanguage [0]{\@gobble}%
\providecommand \bibinfo  [0]{\@secondoftwo}%
\providecommand \bibfield  [0]{\@secondoftwo}%
\providecommand \translation [1]{[#1]}%
\providecommand \BibitemOpen [0]{}%
\providecommand \bibitemStop [0]{}%
\providecommand \bibitemNoStop [0]{.\EOS\space}%
\providecommand \EOS [0]{\spacefactor3000\relax}%
\providecommand \BibitemShut  [1]{\csname bibitem#1\endcsname}%
\let\auto@bib@innerbib\@empty
\bibitem [{\citenamefont {Zhang}\ and\ \citenamefont
  {Averitt}(2014)}]{zhang2014}%
  \BibitemOpen
  \bibfield  {author} {\bibinfo {author} {\bibfnamefont {J.}~\bibnamefont
  {Zhang}}\ and\ \bibinfo {author} {\bibfnamefont {R.}~\bibnamefont
  {Averitt}},\ }\href {\doibase 10.1146/annurev-matsci-070813-113258}
  {\bibfield  {journal} {\bibinfo  {journal} {Annu. Rev. Mater. Res.}\ }\textbf
  {\bibinfo {volume} {44}},\ \bibinfo {pages} {19} (\bibinfo {year}
  {2014})}\BibitemShut {NoStop}%
\bibitem [{\citenamefont {Basov}\ \emph {et~al.}(2017)\citenamefont {Basov},
  \citenamefont {Averitt},\ and\ \citenamefont {Hsieh}}]{basov2017}%
  \BibitemOpen
  \bibfield  {author} {\bibinfo {author} {\bibfnamefont {D.~N.}\ \bibnamefont
  {Basov}}, \bibinfo {author} {\bibfnamefont {R.~D.}\ \bibnamefont {Averitt}},
  \ and\ \bibinfo {author} {\bibfnamefont {D.}~\bibnamefont {Hsieh}},\ }\href
  {\doibase 10.1038/NMAT5017} {\bibfield  {journal} {\bibinfo  {journal} {Nat.
  Mater.}\ }\textbf {\bibinfo {volume} {16}},\ \bibinfo {pages} {1077}
  (\bibinfo {year} {2017})}\BibitemShut {NoStop}%
\bibitem [{\citenamefont {Ishihara}(2019)}]{ishihara2019}%
  \BibitemOpen
  \bibfield  {author} {\bibinfo {author} {\bibfnamefont {S.}~\bibnamefont
  {Ishihara}},\ }\href {\doibase 10.7566/JPSJ.88.072001} {\bibfield  {journal}
  {\bibinfo  {journal} {J. Phys. Soc. Jpn.}\ }\textbf {\bibinfo {volume}
  {88}},\ \bibinfo {pages} {072001} (\bibinfo {year} {2019})}\BibitemShut
  {NoStop}%
\bibitem [{\citenamefont {Giannetti}\ \emph {et~al.}(2016)\citenamefont
  {Giannetti}, \citenamefont {Capone}, \citenamefont {Fausti}, \citenamefont
  {Fabrizio}, \citenamefont {Parmigiani},\ and\ \citenamefont
  {Mihailovic}}]{giannetti2016}%
  \BibitemOpen
  \bibfield  {author} {\bibinfo {author} {\bibfnamefont {C.}~\bibnamefont
  {Giannetti}}, \bibinfo {author} {\bibfnamefont {M.}~\bibnamefont {Capone}},
  \bibinfo {author} {\bibfnamefont {D.}~\bibnamefont {Fausti}}, \bibinfo
  {author} {\bibfnamefont {M.}~\bibnamefont {Fabrizio}}, \bibinfo {author}
  {\bibfnamefont {F.}~\bibnamefont {Parmigiani}}, \ and\ \bibinfo {author}
  {\bibfnamefont {D.}~\bibnamefont {Mihailovic}},\ }\href {\doibase
  10.1080/00018732.2016.1194044} {\bibfield  {journal} {\bibinfo  {journal}
  {Adv. Phys.}\ }\textbf {\bibinfo {volume} {65}},\ \bibinfo {pages} {58}
  (\bibinfo {year} {2016})}\BibitemShut {NoStop}%
\bibitem [{\citenamefont {Fausti}\ \emph {et~al.}(2011)\citenamefont {Fausti},
  \citenamefont {Tobey}, \citenamefont {Dean}, \citenamefont {Kaiser},
  \citenamefont {Dienst}, \citenamefont {Hoffmann}, \citenamefont {Pyon},
  \citenamefont {Takayama}, \citenamefont {Takagi},\ and\ \citenamefont
  {Cavalleri}}]{fausti2011}%
  \BibitemOpen
  \bibfield  {author} {\bibinfo {author} {\bibfnamefont {D.}~\bibnamefont
  {Fausti}}, \bibinfo {author} {\bibfnamefont {R.~I.}\ \bibnamefont {Tobey}},
  \bibinfo {author} {\bibfnamefont {N.}~\bibnamefont {Dean}}, \bibinfo {author}
  {\bibfnamefont {S.}~\bibnamefont {Kaiser}}, \bibinfo {author} {\bibfnamefont
  {A.}~\bibnamefont {Dienst}}, \bibinfo {author} {\bibfnamefont {M.~C.}\
  \bibnamefont {Hoffmann}}, \bibinfo {author} {\bibfnamefont {S.}~\bibnamefont
  {Pyon}}, \bibinfo {author} {\bibfnamefont {T.}~\bibnamefont {Takayama}},
  \bibinfo {author} {\bibfnamefont {H.}~\bibnamefont {Takagi}}, \ and\ \bibinfo
  {author} {\bibfnamefont {A.}~\bibnamefont {Cavalleri}},\ }\href {\doibase
  10.1126/science.1197294} {\bibfield  {journal} {\bibinfo  {journal}
  {Science}\ }\textbf {\bibinfo {volume} {331}},\ \bibinfo {pages} {189}
  (\bibinfo {year} {2011})}\BibitemShut {NoStop}%
\bibitem [{\citenamefont {Hu}\ \emph {et~al.}(2014)\citenamefont {Hu},
  \citenamefont {Kaiser}, \citenamefont {Nicoletti}, \citenamefont {Hunt},
  \citenamefont {Gierz}, \citenamefont {Hoffmann}, \citenamefont {Le~Tacon},
  \citenamefont {Loew}, \citenamefont {Keimer},\ and\ \citenamefont
  {Cavalleri}}]{hu2014}%
  \BibitemOpen
  \bibfield  {author} {\bibinfo {author} {\bibfnamefont {W.}~\bibnamefont
  {Hu}}, \bibinfo {author} {\bibfnamefont {S.}~\bibnamefont {Kaiser}}, \bibinfo
  {author} {\bibfnamefont {D.}~\bibnamefont {Nicoletti}}, \bibinfo {author}
  {\bibfnamefont {C.~R.}\ \bibnamefont {Hunt}}, \bibinfo {author}
  {\bibfnamefont {I.}~\bibnamefont {Gierz}}, \bibinfo {author} {\bibfnamefont
  {M.~C.}\ \bibnamefont {Hoffmann}}, \bibinfo {author} {\bibfnamefont
  {M.}~\bibnamefont {Le~Tacon}}, \bibinfo {author} {\bibfnamefont
  {T.}~\bibnamefont {Loew}}, \bibinfo {author} {\bibfnamefont {B.}~\bibnamefont
  {Keimer}}, \ and\ \bibinfo {author} {\bibfnamefont {A.}~\bibnamefont
  {Cavalleri}},\ }\href {\doibase 10.1038/nmat3963} {\bibfield  {journal}
  {\bibinfo  {journal} {Nat. Mater.}\ }\textbf {\bibinfo {volume} {13}},\
  \bibinfo {pages} {705} (\bibinfo {year} {2014})}\BibitemShut {NoStop}%
\bibitem [{\citenamefont {Kaiser}\ \emph {et~al.}(2014)\citenamefont {Kaiser},
  \citenamefont {Hunt}, \citenamefont {Nicoletti}, \citenamefont {Hu},
  \citenamefont {Gierz}, \citenamefont {Liu}, \citenamefont {Le~Tacon},
  \citenamefont {Loew}, \citenamefont {Haug}, \citenamefont {Keimer},\ and\
  \citenamefont {Cavalleri}}]{kaiser2014}%
  \BibitemOpen
  \bibfield  {author} {\bibinfo {author} {\bibfnamefont {S.}~\bibnamefont
  {Kaiser}}, \bibinfo {author} {\bibfnamefont {C.~R.}\ \bibnamefont {Hunt}},
  \bibinfo {author} {\bibfnamefont {D.}~\bibnamefont {Nicoletti}}, \bibinfo
  {author} {\bibfnamefont {W.}~\bibnamefont {Hu}}, \bibinfo {author}
  {\bibfnamefont {I.}~\bibnamefont {Gierz}}, \bibinfo {author} {\bibfnamefont
  {H.~Y.}\ \bibnamefont {Liu}}, \bibinfo {author} {\bibfnamefont
  {M.}~\bibnamefont {Le~Tacon}}, \bibinfo {author} {\bibfnamefont
  {T.}~\bibnamefont {Loew}}, \bibinfo {author} {\bibfnamefont {D.}~\bibnamefont
  {Haug}}, \bibinfo {author} {\bibfnamefont {B.}~\bibnamefont {Keimer}}, \ and\
  \bibinfo {author} {\bibfnamefont {A.}~\bibnamefont {Cavalleri}},\ }\href
  {\doibase 10.1103/PhysRevB.89.184516} {\bibfield  {journal} {\bibinfo
  {journal} {Phys. Rev. B}\ }\textbf {\bibinfo {volume} {89}},\ \bibinfo
  {pages} {184516} (\bibinfo {year} {2014})}\BibitemShut {NoStop}%
\bibitem [{\citenamefont {Nicoletti}\ \emph {et~al.}(2014)\citenamefont
  {Nicoletti}, \citenamefont {Casandruc}, \citenamefont {Laplace},
  \citenamefont {Khanna}, \citenamefont {Hunt}, \citenamefont {Kaiser},
  \citenamefont {Dhesi}, \citenamefont {Gu}, \citenamefont {Hill},\ and\
  \citenamefont {Cavalleri}}]{nicoletti2014}%
  \BibitemOpen
  \bibfield  {author} {\bibinfo {author} {\bibfnamefont {D.}~\bibnamefont
  {Nicoletti}}, \bibinfo {author} {\bibfnamefont {E.}~\bibnamefont
  {Casandruc}}, \bibinfo {author} {\bibfnamefont {Y.}~\bibnamefont {Laplace}},
  \bibinfo {author} {\bibfnamefont {V.}~\bibnamefont {Khanna}}, \bibinfo
  {author} {\bibfnamefont {C.~R.}\ \bibnamefont {Hunt}}, \bibinfo {author}
  {\bibfnamefont {S.}~\bibnamefont {Kaiser}}, \bibinfo {author} {\bibfnamefont
  {S.~S.}\ \bibnamefont {Dhesi}}, \bibinfo {author} {\bibfnamefont {G.~D.}\
  \bibnamefont {Gu}}, \bibinfo {author} {\bibfnamefont {J.~P.}\ \bibnamefont
  {Hill}}, \ and\ \bibinfo {author} {\bibfnamefont {A.}~\bibnamefont
  {Cavalleri}},\ }\href {\doibase 10.1103/PhysRevB.90.100503} {\bibfield
  {journal} {\bibinfo  {journal} {Phys. Rev. B}\ }\textbf {\bibinfo {volume}
  {90}},\ \bibinfo {pages} {100503(R)} (\bibinfo {year} {2014})}\BibitemShut
  {NoStop}%
\bibitem [{\citenamefont {Mitrano}\ \emph {et~al.}(2016)\citenamefont
  {Mitrano}, \citenamefont {Cantaluppi}, \citenamefont {Nicoletti},
  \citenamefont {Kaiser}, \citenamefont {Perucchi}, \citenamefont {Lupi},
  \citenamefont {Di~Pietro}, \citenamefont {Pontiroli}, \citenamefont
  {Ricc{\`o}}, \citenamefont {Clark}, \citenamefont {Jaksch},\ and\
  \citenamefont {Cavalleri}}]{mitrano2016}%
  \BibitemOpen
  \bibfield  {author} {\bibinfo {author} {\bibfnamefont {M.}~\bibnamefont
  {Mitrano}}, \bibinfo {author} {\bibfnamefont {A.}~\bibnamefont {Cantaluppi}},
  \bibinfo {author} {\bibfnamefont {D.}~\bibnamefont {Nicoletti}}, \bibinfo
  {author} {\bibfnamefont {S.}~\bibnamefont {Kaiser}}, \bibinfo {author}
  {\bibfnamefont {A.}~\bibnamefont {Perucchi}}, \bibinfo {author}
  {\bibfnamefont {S.}~\bibnamefont {Lupi}}, \bibinfo {author} {\bibfnamefont
  {P.}~\bibnamefont {Di~Pietro}}, \bibinfo {author} {\bibfnamefont
  {D.}~\bibnamefont {Pontiroli}}, \bibinfo {author} {\bibfnamefont
  {M.}~\bibnamefont {Ricc{\`o}}}, \bibinfo {author} {\bibfnamefont {S.~R.}\
  \bibnamefont {Clark}}, \bibinfo {author} {\bibfnamefont {D.}~\bibnamefont
  {Jaksch}}, \ and\ \bibinfo {author} {\bibfnamefont {A.}~\bibnamefont
  {Cavalleri}},\ }\href {\doibase 10.1038/nature16522} {\bibfield  {journal}
  {\bibinfo  {journal} {Nature (London)}\ }\textbf {\bibinfo {volume} {530}},\
  \bibinfo {pages} {461} (\bibinfo {year} {2016})}\BibitemShut {NoStop}%
\bibitem [{\citenamefont {Cantaluppi}\ \emph {et~al.}(2018)\citenamefont
  {Cantaluppi}, \citenamefont {Buzzi}, \citenamefont {Jotzu}, \citenamefont
  {Nicoletti}, \citenamefont {Mitrano}, \citenamefont {Pontiroli},
  \citenamefont {Ricc{\`o}}, \citenamefont {Perucchi}, \citenamefont
  {Di~Pietro},\ and\ \citenamefont {Cavalleri}}]{cantaluppi2018}%
  \BibitemOpen
  \bibfield  {author} {\bibinfo {author} {\bibfnamefont {A.}~\bibnamefont
  {Cantaluppi}}, \bibinfo {author} {\bibfnamefont {M.}~\bibnamefont {Buzzi}},
  \bibinfo {author} {\bibfnamefont {G.}~\bibnamefont {Jotzu}}, \bibinfo
  {author} {\bibfnamefont {D.}~\bibnamefont {Nicoletti}}, \bibinfo {author}
  {\bibfnamefont {M.}~\bibnamefont {Mitrano}}, \bibinfo {author} {\bibfnamefont
  {D.}~\bibnamefont {Pontiroli}}, \bibinfo {author} {\bibfnamefont
  {M.}~\bibnamefont {Ricc{\`o}}}, \bibinfo {author} {\bibfnamefont
  {A.}~\bibnamefont {Perucchi}}, \bibinfo {author} {\bibfnamefont
  {P.}~\bibnamefont {Di~Pietro}}, \ and\ \bibinfo {author} {\bibfnamefont
  {A.}~\bibnamefont {Cavalleri}},\ }\href {\doibase 10.1038/s41567-018-0134-8}
  {\bibfield  {journal} {\bibinfo  {journal} {Nat. Phys.}\ }\textbf {\bibinfo
  {volume} {14}},\ \bibinfo {pages} {837} (\bibinfo {year} {2018})}\BibitemShut
  {NoStop}%
\bibitem [{\citenamefont {Sentef}\ \emph {et~al.}(2016)\citenamefont {Sentef},
  \citenamefont {Kemper}, \citenamefont {Georges},\ and\ \citenamefont
  {Kollath}}]{sentef2016}%
  \BibitemOpen
  \bibfield  {author} {\bibinfo {author} {\bibfnamefont {M.~A.}\ \bibnamefont
  {Sentef}}, \bibinfo {author} {\bibfnamefont {A.~F.}\ \bibnamefont {Kemper}},
  \bibinfo {author} {\bibfnamefont {A.}~\bibnamefont {Georges}}, \ and\
  \bibinfo {author} {\bibfnamefont {C.}~\bibnamefont {Kollath}},\ }\href
  {\doibase 10.1103/PhysRevB.93.144506} {\bibfield  {journal} {\bibinfo
  {journal} {Phys. Rev. B}\ }\textbf {\bibinfo {volume} {93}},\ \bibinfo
  {pages} {144506} (\bibinfo {year} {2016})}\BibitemShut {NoStop}%
\bibitem [{\citenamefont {Patel}\ and\ \citenamefont
  {Eberlein}(2016)}]{patel2016}%
  \BibitemOpen
  \bibfield  {author} {\bibinfo {author} {\bibfnamefont {A.~A.}\ \bibnamefont
  {Patel}}\ and\ \bibinfo {author} {\bibfnamefont {A.}~\bibnamefont
  {Eberlein}},\ }\href {\doibase 10.1103/PhysRevB.93.195139} {\bibfield
  {journal} {\bibinfo  {journal} {Phys. Rev. B}\ }\textbf {\bibinfo {volume}
  {93}},\ \bibinfo {pages} {195139} (\bibinfo {year} {2016})}\BibitemShut
  {NoStop}%
\bibitem [{\citenamefont {Knap}\ \emph {et~al.}(2016)\citenamefont {Knap},
  \citenamefont {Babadi}, \citenamefont {Refael}, \citenamefont {Martin},\ and\
  \citenamefont {Demler}}]{knap2016}%
  \BibitemOpen
  \bibfield  {author} {\bibinfo {author} {\bibfnamefont {M.}~\bibnamefont
  {Knap}}, \bibinfo {author} {\bibfnamefont {M.}~\bibnamefont {Babadi}},
  \bibinfo {author} {\bibfnamefont {G.}~\bibnamefont {Refael}}, \bibinfo
  {author} {\bibfnamefont {I.}~\bibnamefont {Martin}}, \ and\ \bibinfo {author}
  {\bibfnamefont {E.}~\bibnamefont {Demler}},\ }\href {\doibase
  10.1103/PhysRevB.94.214504} {\bibfield  {journal} {\bibinfo  {journal} {Phys.
  Rev. B}\ }\textbf {\bibinfo {volume} {94}},\ \bibinfo {pages} {214504}
  (\bibinfo {year} {2016})}\BibitemShut {NoStop}%
\bibitem [{\citenamefont {Kennes}\ \emph {et~al.}(2017)\citenamefont {Kennes},
  \citenamefont {Wilner}, \citenamefont {Reichman},\ and\ \citenamefont
  {Millis}}]{kennes2017}%
  \BibitemOpen
  \bibfield  {author} {\bibinfo {author} {\bibfnamefont {D.~M.}\ \bibnamefont
  {Kennes}}, \bibinfo {author} {\bibfnamefont {E.~Y.}\ \bibnamefont {Wilner}},
  \bibinfo {author} {\bibfnamefont {D.~R.}\ \bibnamefont {Reichman}}, \ and\
  \bibinfo {author} {\bibfnamefont {A.~J.}\ \bibnamefont {Millis}},\ }\href
  {\doibase 10.1038/nphys4024} {\bibfield  {journal} {\bibinfo  {journal} {Nat.
  Phys.}\ }\textbf {\bibinfo {volume} {13}},\ \bibinfo {pages} {479} (\bibinfo
  {year} {2017})}\BibitemShut {NoStop}%
\bibitem [{\citenamefont {Sentef}(2017)}]{sentef2017}%
  \BibitemOpen
  \bibfield  {author} {\bibinfo {author} {\bibfnamefont {M.~A.}\ \bibnamefont
  {Sentef}},\ }\href {\doibase 10.1103/PhysRevB.95.205111} {\bibfield
  {journal} {\bibinfo  {journal} {Phys. Rev. B}\ }\textbf {\bibinfo {volume}
  {95}},\ \bibinfo {pages} {205111} (\bibinfo {year} {2017})}\BibitemShut
  {NoStop}%
\bibitem [{\citenamefont {Babadi}\ \emph {et~al.}(2017)\citenamefont {Babadi},
  \citenamefont {Knap}, \citenamefont {Martin}, \citenamefont {Refael},\ and\
  \citenamefont {Demler}}]{babadi2017}%
  \BibitemOpen
  \bibfield  {author} {\bibinfo {author} {\bibfnamefont {M.}~\bibnamefont
  {Babadi}}, \bibinfo {author} {\bibfnamefont {M.}~\bibnamefont {Knap}},
  \bibinfo {author} {\bibfnamefont {I.}~\bibnamefont {Martin}}, \bibinfo
  {author} {\bibfnamefont {G.}~\bibnamefont {Refael}}, \ and\ \bibinfo {author}
  {\bibfnamefont {E.}~\bibnamefont {Demler}},\ }\href {\doibase
  10.1103/PhysRevB.96.014512} {\bibfield  {journal} {\bibinfo  {journal} {Phys.
  Rev. B}\ }\textbf {\bibinfo {volume} {96}},\ \bibinfo {pages} {014512}
  (\bibinfo {year} {2017})}\BibitemShut {NoStop}%
\bibitem [{\citenamefont {Murakami}\ \emph {et~al.}(2017)\citenamefont
  {Murakami}, \citenamefont {Tsuji}, \citenamefont {Eckstein},\ and\
  \citenamefont {Werner}}]{murakami2017}%
  \BibitemOpen
  \bibfield  {author} {\bibinfo {author} {\bibfnamefont {Y.}~\bibnamefont
  {Murakami}}, \bibinfo {author} {\bibfnamefont {N.}~\bibnamefont {Tsuji}},
  \bibinfo {author} {\bibfnamefont {M.}~\bibnamefont {Eckstein}}, \ and\
  \bibinfo {author} {\bibfnamefont {P.}~\bibnamefont {Werner}},\ }\href
  {\doibase 10.1103/PhysRevB.96.045125} {\bibfield  {journal} {\bibinfo
  {journal} {Phys. Rev. B}\ }\textbf {\bibinfo {volume} {96}},\ \bibinfo
  {pages} {045125} (\bibinfo {year} {2017})}\BibitemShut {NoStop}%
\bibitem [{\citenamefont {Mazza}\ and\ \citenamefont
  {Georges}(2017)}]{mazza2017}%
  \BibitemOpen
  \bibfield  {author} {\bibinfo {author} {\bibfnamefont {G.}~\bibnamefont
  {Mazza}}\ and\ \bibinfo {author} {\bibfnamefont {A.}~\bibnamefont
  {Georges}},\ }\href {\doibase 10.1103/PhysRevB.96.064515} {\bibfield
  {journal} {\bibinfo  {journal} {Phys. Rev. B}\ }\textbf {\bibinfo {volume}
  {96}},\ \bibinfo {pages} {064515} (\bibinfo {year} {2017})}\BibitemShut
  {NoStop}%
\bibitem [{\citenamefont {Wang}\ \emph {et~al.}(2018)\citenamefont {Wang},
  \citenamefont {Chen}, \citenamefont {Moritz},\ and\ \citenamefont
  {Devereaux}}]{wang2018}%
  \BibitemOpen
  \bibfield  {author} {\bibinfo {author} {\bibfnamefont {Y.}~\bibnamefont
  {Wang}}, \bibinfo {author} {\bibfnamefont {C.-C.}\ \bibnamefont {Chen}},
  \bibinfo {author} {\bibfnamefont {B.}~\bibnamefont {Moritz}}, \ and\ \bibinfo
  {author} {\bibfnamefont {T.~P.}\ \bibnamefont {Devereaux}},\ }\href {\doibase
  10.1103/PhysRevLett.120.246402} {\bibfield  {journal} {\bibinfo  {journal}
  {Phys. Rev. Lett.}\ }\textbf {\bibinfo {volume} {120}},\ \bibinfo {pages}
  {246402} (\bibinfo {year} {2018})}\BibitemShut {NoStop}%
\bibitem [{\citenamefont {Floquet}(1883)}]{floquet1883}%
  \BibitemOpen
  \bibfield  {author} {\bibinfo {author} {\bibfnamefont {G.}~\bibnamefont
  {Floquet}},\ }\href {\doibase 10.24033/asens.220} {\bibfield  {journal}
  {\bibinfo  {journal} {Ann. Sci. {\'E}cole Norm. Sup.}\ }\textbf {\bibinfo
  {volume} {12}},\ \bibinfo {pages} {47} (\bibinfo {year} {1883})}\BibitemShut
  {NoStop}%
\bibitem [{\citenamefont {Oka}\ and\ \citenamefont {Kitamura}(2019)}]{oka2019}%
  \BibitemOpen
  \bibfield  {author} {\bibinfo {author} {\bibfnamefont {T.}~\bibnamefont
  {Oka}}\ and\ \bibinfo {author} {\bibfnamefont {S.}~\bibnamefont {Kitamura}},\
  }\href {\doibase 10.1146/annurev-conmatphys-031218-013423} {\bibfield
  {journal} {\bibinfo  {journal} {Annu. Rev. Condens. Matter Phys.}\ }\textbf
  {\bibinfo {volume} {10}},\ \bibinfo {pages} {387} (\bibinfo {year}
  {2019})}\BibitemShut {NoStop}%
\bibitem [{\citenamefont {Micnas}\ \emph {et~al.}(1990)\citenamefont {Micnas},
  \citenamefont {Ranninger},\ and\ \citenamefont {Robaszkiewicz}}]{micnas1990}%
  \BibitemOpen
  \bibfield  {author} {\bibinfo {author} {\bibfnamefont {R.}~\bibnamefont
  {Micnas}}, \bibinfo {author} {\bibfnamefont {J.}~\bibnamefont {Ranninger}}, \
  and\ \bibinfo {author} {\bibfnamefont {S.}~\bibnamefont {Robaszkiewicz}},\
  }\href {\doibase 10.1103/RevModPhys.62.113} {\bibfield  {journal} {\bibinfo
  {journal} {Rev. Mod. Phys.}\ }\textbf {\bibinfo {volume} {62}},\ \bibinfo
  {pages} {113} (\bibinfo {year} {1990})}\BibitemShut {NoStop}%
\bibitem [{\citenamefont {Sentef}\ \emph {et~al.}(2017)\citenamefont {Sentef},
  \citenamefont {Tokuno}, \citenamefont {Georges},\ and\ \citenamefont
  {Kollath}}]{sentef2017a}%
  \BibitemOpen
  \bibfield  {author} {\bibinfo {author} {\bibfnamefont {M.~A.}\ \bibnamefont
  {Sentef}}, \bibinfo {author} {\bibfnamefont {A.}~\bibnamefont {Tokuno}},
  \bibinfo {author} {\bibfnamefont {A.}~\bibnamefont {Georges}}, \ and\
  \bibinfo {author} {\bibfnamefont {C.}~\bibnamefont {Kollath}},\ }\href
  {\doibase 10.1103/PhysRevLett.118.087002} {\bibfield  {journal} {\bibinfo
  {journal} {Phys. Rev. Lett.}\ }\textbf {\bibinfo {volume} {118}},\ \bibinfo
  {pages} {087002} (\bibinfo {year} {2017})}\BibitemShut {NoStop}%
\bibitem [{\citenamefont {Kitamura}\ and\ \citenamefont
  {Aoki}(2016)}]{kitamura2016}%
  \BibitemOpen
  \bibfield  {author} {\bibinfo {author} {\bibfnamefont {S.}~\bibnamefont
  {Kitamura}}\ and\ \bibinfo {author} {\bibfnamefont {H.}~\bibnamefont
  {Aoki}},\ }\href {\doibase 10.1103/PhysRevB.94.174503} {\bibfield  {journal}
  {\bibinfo  {journal} {Phys. Rev. B}\ }\textbf {\bibinfo {volume} {94}},\
  \bibinfo {pages} {174503} (\bibinfo {year} {2016})}\BibitemShut {NoStop}%
\bibitem [{\citenamefont {Yang}(1989)}]{yang1989}%
  \BibitemOpen
  \bibfield  {author} {\bibinfo {author} {\bibfnamefont {C.~N.}\ \bibnamefont
  {Yang}},\ }\href {\doibase 10.1103/PhysRevLett.63.2144} {\bibfield  {journal}
  {\bibinfo  {journal} {Phys. Rev. Lett.}\ }\textbf {\bibinfo {volume} {63}},\
  \bibinfo {pages} {2144} (\bibinfo {year} {1989})}\BibitemShut {NoStop}%
\bibitem [{\citenamefont {Rosch}\ \emph {et~al.}(2008)\citenamefont {Rosch},
  \citenamefont {Rasch}, \citenamefont {Binz},\ and\ \citenamefont
  {Vojta}}]{rosch2008}%
  \BibitemOpen
  \bibfield  {author} {\bibinfo {author} {\bibfnamefont {A.}~\bibnamefont
  {Rosch}}, \bibinfo {author} {\bibfnamefont {D.}~\bibnamefont {Rasch}},
  \bibinfo {author} {\bibfnamefont {B.}~\bibnamefont {Binz}}, \ and\ \bibinfo
  {author} {\bibfnamefont {M.}~\bibnamefont {Vojta}},\ }\href {\doibase
  10.1103/PhysRevLett.101.265301} {\bibfield  {journal} {\bibinfo  {journal}
  {Phys. Rev. Lett.}\ }\textbf {\bibinfo {volume} {101}},\ \bibinfo {pages}
  {265301} (\bibinfo {year} {2008})}\BibitemShut {NoStop}%
\bibitem [{\citenamefont {Yang}\ and\ \citenamefont {Zhang}(1990)}]{yang1990}%
  \BibitemOpen
  \bibfield  {author} {\bibinfo {author} {\bibfnamefont {C.~N.}\ \bibnamefont
  {Yang}}\ and\ \bibinfo {author} {\bibfnamefont {S.}~\bibnamefont {Zhang}},\
  }\href {\doibase 10.1142/S0217984990000933} {\bibfield  {journal} {\bibinfo
  {journal} {Mod. Phys. Lett. B}\ }\textbf {\bibinfo {volume} {04}},\ \bibinfo
  {pages} {759} (\bibinfo {year} {1990})}\BibitemShut {NoStop}%
\bibitem [{\citenamefont {Essler}\ \emph {et~al.}(2005)\citenamefont {Essler},
  \citenamefont {Frahm}, \citenamefont {G{\"o}hmann}, \citenamefont
  {Kl{\"u}mper},\ and\ \citenamefont {Korepin}}]{essler2005}%
  \BibitemOpen
  \bibfield  {author} {\bibinfo {author} {\bibfnamefont {F.~H.}\ \bibnamefont
  {Essler}}, \bibinfo {author} {\bibfnamefont {H.}~\bibnamefont {Frahm}},
  \bibinfo {author} {\bibfnamefont {F.}~\bibnamefont {G{\"o}hmann}}, \bibinfo
  {author} {\bibfnamefont {A.}~\bibnamefont {Kl{\"u}mper}}, \ and\ \bibinfo
  {author} {\bibfnamefont {V.~E.}\ \bibnamefont {Korepin}},\ }\href@noop {}
  {\emph {\bibinfo {title} {The One-Dimensional Hubbard Model}}}\ (\bibinfo
  {publisher} {Cambridge University Press},\ \bibinfo {address} {Cambridge},\
  \bibinfo {year} {2005})\BibitemShut {NoStop}%
\bibitem [{\citenamefont {Shiba}(1972)}]{shiba1972}%
  \BibitemOpen
  \bibfield  {author} {\bibinfo {author} {\bibfnamefont {H.}~\bibnamefont
  {Shiba}},\ }\href {\doibase 10.1143/PTP.48.2171} {\bibfield  {journal}
  {\bibinfo  {journal} {Prog Theor Phys}\ }\textbf {\bibinfo {volume} {48}},\
  \bibinfo {pages} {2171} (\bibinfo {year} {1972})}\BibitemShut {NoStop}%
\bibitem [{\citenamefont {Emery}(1976)}]{emery1976}%
  \BibitemOpen
  \bibfield  {author} {\bibinfo {author} {\bibfnamefont {V.~J.}\ \bibnamefont
  {Emery}},\ }\href {\doibase 10.1103/PhysRevB.14.2989} {\bibfield  {journal}
  {\bibinfo  {journal} {Phys. Rev. B}\ }\textbf {\bibinfo {volume} {14}},\
  \bibinfo {pages} {2989} (\bibinfo {year} {1976})}\BibitemShut {NoStop}%
\bibitem [{\citenamefont {Ono}\ \emph {et~al.}(2017)\citenamefont {Ono},
  \citenamefont {Hashimoto},\ and\ \citenamefont {Ishihara}}]{ono2017}%
  \BibitemOpen
  \bibfield  {author} {\bibinfo {author} {\bibfnamefont {A.}~\bibnamefont
  {Ono}}, \bibinfo {author} {\bibfnamefont {H.}~\bibnamefont {Hashimoto}}, \
  and\ \bibinfo {author} {\bibfnamefont {S.}~\bibnamefont {Ishihara}},\ }\href
  {\doibase 10.1103/PhysRevB.95.085123} {\bibfield  {journal} {\bibinfo
  {journal} {Phys. Rev. B}\ }\textbf {\bibinfo {volume} {95}},\ \bibinfo
  {pages} {085123} (\bibinfo {year} {2017})}\BibitemShut {NoStop}%
\bibitem [{DMR()}]{DMRG}%
  \BibitemOpen
  \href@noop {} {}\bibinfo {note} {Under the strong external field, the
  time-dependent density-matrix renormalization group method requires a large
  amount of computational resource for accurate long-timescale calculations in
  larger systems~\cite{ejima2020}. Therefore, in this study we employ the
  time-dependent ED method.}\BibitemShut {Stop}%
\bibitem [{\citenamefont {Mohankumar}\ and\ \citenamefont
  {Auerbach}(2006)}]{Mohankumar2006}%
  \BibitemOpen
  \bibfield  {author} {\bibinfo {author} {\bibfnamefont {N.}~\bibnamefont
  {Mohankumar}}\ and\ \bibinfo {author} {\bibfnamefont {S.~M.}\ \bibnamefont
  {Auerbach}},\ }\href {\doibase http://dx.doi.org/10.1016/j,\.cpc.2006.07.005}
  {\bibfield  {journal} {\bibinfo  {journal} {Comput. Phys. Commun.}\ }\textbf
  {\bibinfo {volume} {175}},\ \bibinfo {pages} {473} (\bibinfo {year}
  {2006})}\BibitemShut {NoStop}%
\bibitem [{\citenamefont {Park}\ and\ \citenamefont {Light}(1986)}]{Park1986}%
  \BibitemOpen
  \bibfield  {author} {\bibinfo {author} {\bibfnamefont {T.~J.}\ \bibnamefont
  {Park}}\ and\ \bibinfo {author} {\bibfnamefont {J.}~\bibnamefont {Light}},\
  }\href {\doibase 10.1063/1.451548} {\bibfield  {journal} {\bibinfo  {journal}
  {J. Chem. Phys.}\ }\textbf {\bibinfo {volume} {85}},\ \bibinfo {pages} {5870}
  (\bibinfo {year} {1986})}\BibitemShut {NoStop}%
\bibitem [{\citenamefont {Kaneko}\ \emph {et~al.}(2019)\citenamefont {Kaneko},
  \citenamefont {Shirakawa}, \citenamefont {Sorella},\ and\ \citenamefont
  {Yunoki}}]{kaneko2019}%
  \BibitemOpen
  \bibfield  {author} {\bibinfo {author} {\bibfnamefont {T.}~\bibnamefont
  {Kaneko}}, \bibinfo {author} {\bibfnamefont {T.}~\bibnamefont {Shirakawa}},
  \bibinfo {author} {\bibfnamefont {S.}~\bibnamefont {Sorella}}, \ and\
  \bibinfo {author} {\bibfnamefont {S.}~\bibnamefont {Yunoki}},\ }\href
  {\doibase 10.1103/PhysRevLett.122.077002} {\bibfield  {journal} {\bibinfo
  {journal} {Phys. Rev. Lett.}\ }\textbf {\bibinfo {volume} {122}},\ \bibinfo
  {pages} {077002} (\bibinfo {year} {2019})}\BibitemShut {NoStop}%
\bibitem [{\citenamefont {Fujiuchi}\ \emph {et~al.}(2019)\citenamefont
  {Fujiuchi}, \citenamefont {Kaneko}, \citenamefont {Ohta},\ and\ \citenamefont
  {Yunoki}}]{fujiuchi2019}%
  \BibitemOpen
  \bibfield  {author} {\bibinfo {author} {\bibfnamefont {R.}~\bibnamefont
  {Fujiuchi}}, \bibinfo {author} {\bibfnamefont {T.}~\bibnamefont {Kaneko}},
  \bibinfo {author} {\bibfnamefont {Y.}~\bibnamefont {Ohta}}, \ and\ \bibinfo
  {author} {\bibfnamefont {S.}~\bibnamefont {Yunoki}},\ }\href {\doibase
  10.1103/PhysRevB.100.045121} {\bibfield  {journal} {\bibinfo  {journal}
  {Phys. Rev. B}\ }\textbf {\bibinfo {volume} {100}},\ \bibinfo {pages}
  {045121} (\bibinfo {year} {2019})}\BibitemShut {NoStop}%
\bibitem [{\citenamefont {Kaneko}\ \emph {et~al.}()\citenamefont {Kaneko},
  \citenamefont {Yunoki},\ and\ \citenamefont {Millis}}]{kaneko2020}%
  \BibitemOpen
  \bibfield  {author} {\bibinfo {author} {\bibfnamefont {T.}~\bibnamefont
  {Kaneko}}, \bibinfo {author} {\bibfnamefont {S.}~\bibnamefont {Yunoki}}, \
  and\ \bibinfo {author} {\bibfnamefont {A.~J.}\ \bibnamefont {Millis}},\
  }\href@noop {} {\ }\Eprint {http://arxiv.org/abs/1910.11229}
  {arXiv:1910.11229} \BibitemShut {NoStop}%
\bibitem [{\citenamefont {Dunlap}\ and\ \citenamefont
  {Kenkre}(1986)}]{dunlap1986}%
  \BibitemOpen
  \bibfield  {author} {\bibinfo {author} {\bibfnamefont {D.~H.}\ \bibnamefont
  {Dunlap}}\ and\ \bibinfo {author} {\bibfnamefont {V.~M.}\ \bibnamefont
  {Kenkre}},\ }\href {\doibase 10.1103/PhysRevB.34.3625} {\bibfield  {journal}
  {\bibinfo  {journal} {Phys. Rev. B}\ }\textbf {\bibinfo {volume} {34}},\
  \bibinfo {pages} {3625} (\bibinfo {year} {1986})}\BibitemShut {NoStop}%
\bibitem [{\citenamefont {Messer}\ \emph {et~al.}(2018)\citenamefont {Messer},
  \citenamefont {Sandholzer}, \citenamefont {G\"org}, \citenamefont {Minguzzi},
  \citenamefont {Desbuquois},\ and\ \citenamefont {Esslinger}}]{messer2018}%
  \BibitemOpen
  \bibfield  {author} {\bibinfo {author} {\bibfnamefont {M.}~\bibnamefont
  {Messer}}, \bibinfo {author} {\bibfnamefont {K.}~\bibnamefont {Sandholzer}},
  \bibinfo {author} {\bibfnamefont {F.}~\bibnamefont {G\"org}}, \bibinfo
  {author} {\bibfnamefont {J.}~\bibnamefont {Minguzzi}}, \bibinfo {author}
  {\bibfnamefont {R.}~\bibnamefont {Desbuquois}}, \ and\ \bibinfo {author}
  {\bibfnamefont {T.}~\bibnamefont {Esslinger}},\ }\href {\doibase
  10.1103/PhysRevLett.121.233603} {\bibfield  {journal} {\bibinfo  {journal}
  {Phys. Rev. Lett.}\ }\textbf {\bibinfo {volume} {121}},\ \bibinfo {pages}
  {233603} (\bibinfo {year} {2018})}\BibitemShut {NoStop}%
\bibitem [{\citenamefont {Sandholzer}\ \emph {et~al.}(2019)\citenamefont
  {Sandholzer}, \citenamefont {Murakami}, \citenamefont {G\"org}, \citenamefont
  {Minguzzi}, \citenamefont {Messer}, \citenamefont {Desbuquois}, \citenamefont
  {Eckstein}, \citenamefont {Werner},\ and\ \citenamefont
  {Esslinger}}]{sandholzer2020}%
  \BibitemOpen
  \bibfield  {author} {\bibinfo {author} {\bibfnamefont {K.}~\bibnamefont
  {Sandholzer}}, \bibinfo {author} {\bibfnamefont {Y.}~\bibnamefont
  {Murakami}}, \bibinfo {author} {\bibfnamefont {F.}~\bibnamefont {G\"org}},
  \bibinfo {author} {\bibfnamefont {J.}~\bibnamefont {Minguzzi}}, \bibinfo
  {author} {\bibfnamefont {M.}~\bibnamefont {Messer}}, \bibinfo {author}
  {\bibfnamefont {R.}~\bibnamefont {Desbuquois}}, \bibinfo {author}
  {\bibfnamefont {M.}~\bibnamefont {Eckstein}}, \bibinfo {author}
  {\bibfnamefont {P.}~\bibnamefont {Werner}}, \ and\ \bibinfo {author}
  {\bibfnamefont {T.}~\bibnamefont {Esslinger}},\ }\href {\doibase
  10.1103/PhysRevLett.123.193602} {\bibfield  {journal} {\bibinfo  {journal}
  {Phys. Rev. Lett.}\ }\textbf {\bibinfo {volume} {123}},\ \bibinfo {pages}
  {193602} (\bibinfo {year} {2019})}\BibitemShut {NoStop}%
\bibitem [{\citenamefont {Mentink}\ \emph {et~al.}(2015)\citenamefont
  {Mentink}, \citenamefont {Balzer},\ and\ \citenamefont
  {Eckstein}}]{mentink2015}%
  \BibitemOpen
  \bibfield  {author} {\bibinfo {author} {\bibfnamefont {J.~H.}\ \bibnamefont
  {Mentink}}, \bibinfo {author} {\bibfnamefont {K.}~\bibnamefont {Balzer}}, \
  and\ \bibinfo {author} {\bibfnamefont {M.}~\bibnamefont {Eckstein}},\ }\href
  {\doibase 10.1038/ncomms7708} {\bibfield  {journal} {\bibinfo  {journal}
  {Nat. Commun.}\ }\textbf {\bibinfo {volume} {6}},\ \bibinfo {pages} {6708}
  (\bibinfo {year} {2015})}\BibitemShut {NoStop}%
\bibitem [{\citenamefont {Mitra}\ \emph {et~al.}(2018)\citenamefont {Mitra},
  \citenamefont {Brown}, \citenamefont {Guardado-Sanchez}, \citenamefont
  {Kondov}, \citenamefont {Devakul}, \citenamefont {Huse}, \citenamefont
  {Schau{\ss}},\ and\ \citenamefont {Bakr}}]{mitra2018}%
  \BibitemOpen
  \bibfield  {author} {\bibinfo {author} {\bibfnamefont {D.}~\bibnamefont
  {Mitra}}, \bibinfo {author} {\bibfnamefont {P.~T.}\ \bibnamefont {Brown}},
  \bibinfo {author} {\bibfnamefont {E.}~\bibnamefont {Guardado-Sanchez}},
  \bibinfo {author} {\bibfnamefont {S.~S.}\ \bibnamefont {Kondov}}, \bibinfo
  {author} {\bibfnamefont {T.}~\bibnamefont {Devakul}}, \bibinfo {author}
  {\bibfnamefont {D.~A.}\ \bibnamefont {Huse}}, \bibinfo {author}
  {\bibfnamefont {P.}~\bibnamefont {Schau{\ss}}}, \ and\ \bibinfo {author}
  {\bibfnamefont {W.~S.}\ \bibnamefont {Bakr}},\ }\href {\doibase
  10.1038/nphys4297} {\bibfield  {journal} {\bibinfo  {journal} {Nat. Phys.}\
  }\textbf {\bibinfo {volume} {14}},\ \bibinfo {pages} {173} (\bibinfo {year}
  {2018})}\BibitemShut {NoStop}%
\bibitem [{\citenamefont {Ghiringhelli}\ \emph {et~al.}(2012)\citenamefont
  {Ghiringhelli}, \citenamefont {Le~Tacon}, \citenamefont {Minola},
  \citenamefont {Blanco-Canosa}, \citenamefont {Mazzoli}, \citenamefont
  {Brookes}, \citenamefont {De~Luca}, \citenamefont {Frano}, \citenamefont
  {Hawthorn}, \citenamefont {He}, \citenamefont {Loew}, \citenamefont {Sala},
  \citenamefont {Peets}, \citenamefont {Salluzzo}, \citenamefont {Schierle},
  \citenamefont {Sutarto}, \citenamefont {Sawatzky}, \citenamefont {Weschke},
  \citenamefont {Keimer},\ and\ \citenamefont {Braicovich}}]{ghiringhelli2012}%
  \BibitemOpen
  \bibfield  {author} {\bibinfo {author} {\bibfnamefont {G.}~\bibnamefont
  {Ghiringhelli}}, \bibinfo {author} {\bibfnamefont {M.}~\bibnamefont
  {Le~Tacon}}, \bibinfo {author} {\bibfnamefont {M.}~\bibnamefont {Minola}},
  \bibinfo {author} {\bibfnamefont {S.}~\bibnamefont {Blanco-Canosa}}, \bibinfo
  {author} {\bibfnamefont {C.}~\bibnamefont {Mazzoli}}, \bibinfo {author}
  {\bibfnamefont {N.~B.}\ \bibnamefont {Brookes}}, \bibinfo {author}
  {\bibfnamefont {G.~M.}\ \bibnamefont {De~Luca}}, \bibinfo {author}
  {\bibfnamefont {A.}~\bibnamefont {Frano}}, \bibinfo {author} {\bibfnamefont
  {D.~G.}\ \bibnamefont {Hawthorn}}, \bibinfo {author} {\bibfnamefont
  {F.}~\bibnamefont {He}}, \bibinfo {author} {\bibfnamefont {T.}~\bibnamefont
  {Loew}}, \bibinfo {author} {\bibfnamefont {M.~M.}\ \bibnamefont {Sala}},
  \bibinfo {author} {\bibfnamefont {D.~C.}\ \bibnamefont {Peets}}, \bibinfo
  {author} {\bibfnamefont {M.}~\bibnamefont {Salluzzo}}, \bibinfo {author}
  {\bibfnamefont {E.}~\bibnamefont {Schierle}}, \bibinfo {author}
  {\bibfnamefont {R.}~\bibnamefont {Sutarto}}, \bibinfo {author} {\bibfnamefont
  {G.~A.}\ \bibnamefont {Sawatzky}}, \bibinfo {author} {\bibfnamefont
  {E.}~\bibnamefont {Weschke}}, \bibinfo {author} {\bibfnamefont
  {B.}~\bibnamefont {Keimer}}, \ and\ \bibinfo {author} {\bibfnamefont
  {L.}~\bibnamefont {Braicovich}},\ }\href {\doibase 10.1126/science.1223532}
  {\bibfield  {journal} {\bibinfo  {journal} {Science}\ }\textbf {\bibinfo
  {volume} {337}},\ \bibinfo {pages} {821} (\bibinfo {year}
  {2012})}\BibitemShut {NoStop}%
\bibitem [{\citenamefont {Chang}\ \emph {et~al.}(2012)\citenamefont {Chang},
  \citenamefont {Blackburn}, \citenamefont {Holmes}, \citenamefont
  {Christensen}, \citenamefont {Larsen}, \citenamefont {Mesot}, \citenamefont
  {Liang}, \citenamefont {Bonn}, \citenamefont {Hardy}, \citenamefont
  {Watenphul}, \citenamefont {Zimmermann}, \citenamefont {Forgan},\ and\
  \citenamefont {Hayden}}]{chang2012}%
  \BibitemOpen
  \bibfield  {author} {\bibinfo {author} {\bibfnamefont {J.}~\bibnamefont
  {Chang}}, \bibinfo {author} {\bibfnamefont {E.}~\bibnamefont {Blackburn}},
  \bibinfo {author} {\bibfnamefont {A.~T.}\ \bibnamefont {Holmes}}, \bibinfo
  {author} {\bibfnamefont {N.~B.}\ \bibnamefont {Christensen}}, \bibinfo
  {author} {\bibfnamefont {J.}~\bibnamefont {Larsen}}, \bibinfo {author}
  {\bibfnamefont {J.}~\bibnamefont {Mesot}}, \bibinfo {author} {\bibfnamefont
  {R.}~\bibnamefont {Liang}}, \bibinfo {author} {\bibfnamefont {D.~A.}\
  \bibnamefont {Bonn}}, \bibinfo {author} {\bibfnamefont {W.~N.}\ \bibnamefont
  {Hardy}}, \bibinfo {author} {\bibfnamefont {A.}~\bibnamefont {Watenphul}},
  \bibinfo {author} {\bibfnamefont {M.~v.}\ \bibnamefont {Zimmermann}},
  \bibinfo {author} {\bibfnamefont {E.~M.}\ \bibnamefont {Forgan}}, \ and\
  \bibinfo {author} {\bibfnamefont {S.~M.}\ \bibnamefont {Hayden}},\ }\href
  {\doibase 10.1038/nphys2456} {\bibfield  {journal} {\bibinfo  {journal} {Nat.
  Phys.}\ }\textbf {\bibinfo {volume} {8}},\ \bibinfo {pages} {871} (\bibinfo
  {year} {2012})}\BibitemShut {NoStop}%
\bibitem [{\citenamefont {Harper}\ \emph {et~al.}(1977)\citenamefont {Harper},
  \citenamefont {Geballe},\ and\ \citenamefont {DiSalvo}}]{harper1977}%
  \BibitemOpen
  \bibfield  {author} {\bibinfo {author} {\bibfnamefont {J.~M.~E.}\
  \bibnamefont {Harper}}, \bibinfo {author} {\bibfnamefont {T.~H.}\
  \bibnamefont {Geballe}}, \ and\ \bibinfo {author} {\bibfnamefont {F.~J.}\
  \bibnamefont {DiSalvo}},\ }\href {\doibase 10.1103/PhysRevB.15.2943}
  {\bibfield  {journal} {\bibinfo  {journal} {Phys. Rev. B}\ }\textbf {\bibinfo
  {volume} {15}},\ \bibinfo {pages} {2943} (\bibinfo {year}
  {1977})}\BibitemShut {NoStop}%
\bibitem [{\citenamefont {Castro~Neto}(2001)}]{neto2001}%
  \BibitemOpen
  \bibfield  {author} {\bibinfo {author} {\bibfnamefont {A.~H.}\ \bibnamefont
  {Castro~Neto}},\ }\href {\doibase 10.1103/PhysRevLett.86.4382} {\bibfield
  {journal} {\bibinfo  {journal} {Phys. Rev. Lett.}\ }\textbf {\bibinfo
  {volume} {86}},\ \bibinfo {pages} {4382} (\bibinfo {year}
  {2001})}\BibitemShut {NoStop}%
\bibitem [{\citenamefont {Freericks}(1993)}]{freericks1993}%
  \BibitemOpen
  \bibfield  {author} {\bibinfo {author} {\bibfnamefont {J.~K.}\ \bibnamefont
  {Freericks}},\ }\href {\doibase 10.1103/PhysRevB.48.3881} {\bibfield
  {journal} {\bibinfo  {journal} {Phys. Rev. B}\ }\textbf {\bibinfo {volume}
  {48}},\ \bibinfo {pages} {3881} (\bibinfo {year} {1993})}\BibitemShut
  {NoStop}%
\bibitem [{\citenamefont {Ejima}\ \emph {et~al.}(2020)\citenamefont {Ejima},
  \citenamefont {Kaneko}, \citenamefont {Lange}, \citenamefont {Yunoki},\ and\
  \citenamefont {Fehske}}]{ejima2020}%
  \BibitemOpen
  \bibfield  {author} {\bibinfo {author} {\bibfnamefont {S.}~\bibnamefont
  {Ejima}}, \bibinfo {author} {\bibfnamefont {T.}~\bibnamefont {Kaneko}},
  \bibinfo {author} {\bibfnamefont {F.}~\bibnamefont {Lange}}, \bibinfo
  {author} {\bibfnamefont {S.}~\bibnamefont {Yunoki}}, \ and\ \bibinfo {author}
  {\bibfnamefont {H.}~\bibnamefont {Fehske}},\ }\href {\doibase
  10.7566/JPSCP.30.011184} {\bibfield  {journal} {\bibinfo  {journal} {JPS
  Conf. Proc.}\ }\textbf {\bibinfo {volume} {30}},\ \bibinfo {pages} {011184}
  (\bibinfo {year} {2020})}\BibitemShut {NoStop}%
\end{thebibliography}%

\end{document}